\documentclass[prd,amsmath,amssymb,showpacs,superscriptaddress,nofootinbib,twocolumn]{revtex4-1}

\usepackage{xcolor}
\usepackage{graphicx}
\usepackage{dcolumn}
\usepackage{bm}
\usepackage[mathlines]{lineno}

\begin{document}

\preprint{APS/123-QED}

\title{Measurements of $e^+e^-\rightarrow \eta_{\rm c}\pi^+ \pi^-\pi^0$, $\eta_{\rm c}\pi^+ \pi^-$ and 
  $\eta_{\rm c}\pi^0\gamma$ at $\sqrt{s}$ from 4.18 to 4.60\,GeV, and search for a
  $Z_{\rm c}$ state close to the $D\bar{D}$ threshold decaying to $\eta_{\rm c}\pi$ at $\sqrt{s}=4.23$\,GeV
}




\author{M.~Ablikim$^{1}$, M.~N.~Achasov$^{10,c}$, P.~Adlarson$^{64}$, S. ~Ahmed$^{15}$, M.~Albrecht$^{4}$, R.~Aliberti$^{28}$, A.~Amoroso$^{63A,63C}$, Q.~An$^{60,48}$, ~Anita$^{21}$, X.~H.~Bai$^{54}$
% ...
\\
\vspace{0.2cm}
(BESIII Collaboration)\\
\vspace{0.2cm} {\it
$^{1}$ Institute of High Energy Physics, Beijing 100049, People's Republic of China\\
% ...
$^{4}$ Bochum Ruhr-University, D-44780 Bochum, Germany\\
% ...
$^{10}$ G.I. Budker Institute of Nuclear Physics SB RAS (BINP), Novosibirsk 630090, Russia\\
% ...
$^{15}$ Helmholtz Institute Mainz, Johann-Joachim-Becher-Weg 45, D-55099 Mainz, Germany\\
% ...
$^{21}$ Indian Institute of Technology Madras, Chennai 600036, India\\
% ...
$^{28}$ Johannes Gutenberg University of Mainz, Johann-Joachim-Becher-Weg 45, D-55099 Mainz, Germany\\
% ...
$^{48}$ State Key Laboratory of Particle Detection and Electronics, Beijing 100049, Hefei 230026, People's Republic of China\\
% ...
$^{54}$ University of Jinan, Jinan 250022, People's Republic of China\\
% ...
$^{60}$ University of Science and Technology of China, Hefei 230026, People's Republic of China\\
% ...
$^{63}$ (A)University of Turin, I-10125, Turin, Italy; (B)University of Eastern Piedmont, I-15121, Alessandria, Italy; (C)INFN, I-10125, Turin, Italy\\
$^{64}$ Uppsala University, Box 516, SE-75120 Uppsala, Sweden\\
% ...
$^{c}$ Also at the Novosibirsk State University, Novosibirsk, 630090, Russia\\
% ...
}
}

\vspace{0.4cm}

\date{\today}

\begin{abstract}
We study $\eta_{\rm c}$ production at center-of-mass energies $\sqrt{s}$ from 4.18 to 4.60\,GeV in $e^+e^-$ annihilation data collected with the BESIII detector operating at the BEPCII storage ring, corresponding to 7.3 fb$^{-1}$ of integrated luminosity. We measure the cross sections of the three different exclusive reactions $e^+e^-\rightarrow \eta_{\rm c}\pi^+ \pi^-\pi^0$, $e^+e^- \rightarrow \eta_{\rm c}\pi^+ \pi^-$, and $e^+e^- \rightarrow \eta_{\rm c}\pi^0\gamma$. We find significant $\eta_{\rm c}$ production in $e^+e^-\rightarrow \eta_{\rm c}\pi^+ \pi^-\pi^0$ at $\sqrt{s}$ of 4.23\,GeV and 4.26\,GeV and observe a significant energy-dependent Born cross section that we measure to be consistent with the production via the intermediate $Y(4260)$ resonance. In addition, we perform a search for a charmonium-like $Z_{\rm c}$ state close to the $D\bar{D}$ threshold that decays to $\eta_{\rm c}\pi$, involving ground state charmonium, and observe no signal. Corresponding upper limits on the cross section of $\eta_{\rm c}$ and $Z_{\rm c}$ production are provided, where the yields are not found to be significant. 

\end{abstract}

\maketitle


\section{\label{sec:Intro} Introduction}
A series of new, unexpected states have been observed at $e^+e^-$ colliders in studies dating back to the beginning of the millenium. As summarized and discussed in recent reviews~\cite{ReviewPapaerMitcheletAl,Olsen:2017bmm,XYZinterpretations,Guo:2017jvc,Esposito:2016noz}, several of these states observed in the charmonium and bottomonium mass regions show characteristics different from predictions for conventional states based on potential models, and they are therefore suggested to be exotic hadron candidates. 

More than 20  states in the charmonium region have been observed in decay modes that indicate an internal structure containing a charm and an anti-charm quark pair.  These resonances are designated as `XYZ states', which signifies them to be charmonium-like.  However their properties, for example their mass or decay patterns, do not allow them to be easily identified as unassigned conventional charmonium states. In order to improve our understanding of the XYZ states, it is important to search for further exotic candidates along with new decay modes of already observed unconventional states as well as new production mechanisms. Prominent examples of the XYZ states are the earliest that were observed: the $X(3872)$~\footnote{The $X(3872)$ has recently been renamed to $\chi_{\rm c1}(3872)$ in tables by the Particle Data Group.}, discovered in 2003~\cite{BelleX3872}, the vector states $Y(4260)$ and $Y(4360)$~\cite{Y4260_1_BaBar,Y4360_1_Belle, Y4360_2_BaBar,Y4xxx_2_Belle}, the $Z_{\rm c}(4430)$~\cite{Zc4430_1_Belle,Zc4430_2_LHCb}, and  $Z_{\rm c}(3900)$~\cite{Zc3900_1_BESIII,Zc3900_2_Belle} states. These resonances have all been observed in decays to final states containing low-mass charmonia, such as $J/\psi\,\pi^+\pi^-$, $\psi^\prime\,\pi^+\pi^-$ and $\psi^\prime\,\pi^\pm$, $J/\psi\,\pi^\pm$, respectively.

The charmonium-like $Z_{\rm c}$ states are of particular interest. Given these are electrically charged, they can not be conventional charmonium states and are thus manifestly exotic. They are mainly considered to be candidates for four-quark configurations, and speculations comprise interpretations such as hadro-charmonium, hadronic molecule, or tetraquark states, see {\it e.g.} Ref.~\cite{ReviewPapaerMitcheletAl}. 

For the $Z_{\rm c}(3900)$ and the $Z_{\rm c}(4020)$~\cite{Zc3900_1_BESIII,Zc3900_2_Belle,Zc4020_1_CLEO,Zc4020_2_BESIII,Zc3885_1_BESIII,Zc4025_1_BESIII} states, neutral isospin partner states have meanwhile been found and established in  BESIII data~\cite{Zc3900_Neutral_BESIII,Zc3885_Neutral_BESIII,Zc4020_Neutral_1_BESIII,Zc4025_Neutral_1_BESIII}, {\it cf.}~\cite{PDG18}. Corresponding spin-parity analyses indicate different observed decay modes (hidden {\it vs.} open charm) of the $Z_{\rm c}(3900)^\pm$ to be decays of the same state~\cite{Zc3885_AD_a_BESIII,Zc3885_AD_b_BESIII,Zc3900_PWA_BESIII}. The quantum numbers for both the charged and the neutral $Z_{\rm c}(3900)$ state have been determined to be $J^{P}= 1^{+}$~\cite{Zc3900_PWA_BESIII,draftJpsipi0pi02020}. 

Despite this remarkable progress, the nature of these states is still unclear. Further decay channels involving, {\it e.g.} $\eta_{\rm c}$, should be searched for to complement those observed into other charmonia states ({\it i.e.} $J/\psi$, $\psi^\prime$ and $h_{\rm c}$), and possible $Z_{\rm c}$ multiplets for spin quantum numbers other than $J^{P}= 1^{+}$ need to be established. 

Interestingly, some of the newly observed states have masses close to open-charm meson pair thresholds, see {\it e.g.}~\cite{ReviewPapaerMitcheletAl}. The mass of the first discovered state, the $X(3872)$, is still experimentally indistinguishable from the $D\bar{D}^\ast$ threshold, and the $Z_{\rm c}(3900)$ and $Z_{\rm c}(4020)$ states are found close to the $D\bar{D}^\ast$ and $D^\ast\bar{D}^\ast$ thresholds, respectively. It is therefore well motivated to search for possible $Z_{\rm c}$-like states close to the $D\bar{D}$ threshold. Given the $D\bar{D}$ threshold is in the mass region of around $3730$ to $3740$\,MeV and the spin-parity of the system $J^P(D\bar{D}) =0^- \,{\small \oplus}\, 0^- = 0^+$, a possible $Z_{\rm c}$ state is expected to decay (with the orbital angular momentum $l=0$)  into ground state charmonium together with another  pseudoscalar, such as $Z_{\rm c}\to \eta_{\rm c}\pi$. Correspondingly, it is  also important to search for decays of vector $Y$ states to the lowest lying charmonium and accompanying light recoil particles, such as $Y(4260) \to \eta_{\rm c} + {\rm light\,recoils}$. 

There are various theoretical models that predict possible $Z_{\rm c}$ states decaying to $\eta_{\rm c}\pi$. The hadro-charmonium model for instance, according to which {\it e.g.} the $Z_{\rm c}(3900)$ is interpreted as a compact $c\bar{c}$ pair loosely bound to the surrounding light quark pair via a QCD analogue of the van der Waals force, predicts a $Z_{\rm c}$-like state of about 3800\,MeV/$c^2$~\cite{VoloshinPRL2013} that would dominantly decay to $\eta_{\rm c}\pi$. Also within the di-quark model, a rich spectrum of hadrons is predicted that comprises the observed exotic states and includes a $J^P=0^+$ candidate just below the $D\bar{D}$ threshold with allowed decays to the $\eta_{\rm c}\pi$ final state~\cite{Maiani_PRD2005}. Based on Lattice QCD, the $Z_{\rm c}$ states can alternatively be interpreted as quarkonium hybrids, leading to predictions of different tetraquark multiplets~\cite{Braaten_PRL2013} that comprise states with $J^{P(C)}$ also allowed to decay to the $\eta_{\rm c}\pi$ final state.

Therefore, the investigation of $e^+e^-$ production of an $\eta_{\rm c}$  meson together with pionic recoil particles will provide important input to the understanding of the nature of the charmonium-like exotic states. In the case of significant $\eta_{\rm c}$ production cross sections, further insight will come from the subsequent search for possible $Z_{\rm c}$-like states close to the open-charm threshold, decaying to $\eta_{\rm c}\pi$, 

First evidence of more than $3\sigma$ significance for a charged charmonium-like $\eta_{\rm c}\pi$ resonance with a mass of $m_{\rm Z_c^-} = (4096 \pm 20^{+18}_{-22})~{\rm MeV}/c^2$ and a width of $\Gamma_{\rm Z_c^-} = (152 \pm 58^{+60}_{-35})~{\rm MeV}$ has meanwhile been reported in $B^0\to K^+ \eta_{\rm c}\pi^-$ decays by the LHCb Collaboration~\cite{etacpi_LHCb_2015}. 
                      
A first study  by {\mbox BESIII} of $e^+e^-\to\eta_{\rm c}\pi^+\pi^-\pi^0$ revealed evidence for $Z_{\rm c}(3900) \to \rho^\pm \eta_{\rm c}$ with a significance of  $3.9\sigma$~\cite{etacRho_BESIII_PRD100_2019_111102}. In this article we report cross section measurements of the three different exclusive $\eta_{\rm c} + {\rm light\,recoil}$ production channels $e^+e^-\rightarrow \eta_{\rm c}\pi^+ \pi^-\pi^0$, $e^+e^- \rightarrow \eta_{\rm c}\pi^+ \pi^-$ and $e^+e^- \rightarrow \eta_{\rm c}\pi^0\gamma$.  In addition, we perform a search for a (charged and neutral) charmonium-like $\eta_{\rm c}\pi$ resonance in $(\eta_{\rm c}\pi^{+})\pi^{-}\pi^{0}$ and $(\eta_{\rm c}\pi^{0})\pi^{+}\pi^{-}$ final states, as here a significant underlying energy-dependent $\eta_{\rm c}$ cross section is observed. 

The paper continues in Sec.\,\ref{sec:DetsData} with a brief description of the BESIII detector, the data samples, as well as the reconstruction and simulation software used. The event selections, determination of reconstruction efficiencies and estimation of radiative corrections are presented  in Sec.\,\ref{sec:Ana}. The cross section measurements, employing a simultaneous fit to 16 hadronic $\eta_{\rm c}$ decay channels for the three different $\eta_{\rm c} + {\rm light\,recoil} $ production channels, are discussed in Sec.\,\ref{sec:Xsec}, and a subsequent search for a $Z_{\rm c}$-like $\eta_{\rm c}\pi$ resonance close to the $D\bar{D}$ threshold   is described in Sec.\,\ref{sec:Zc}. The systematic uncertainties are evaluated in Sec.\ref{sec:Syst}, and the results are finally summarized in Sec.\,\ref{sec:Summary}.

\section{\label{sec:DetsData} Detectors and data samples}
The BESIII experiment~\cite{bes3} at the BEPCII collider~\cite{bpc2} is a general purpose magnetic spectrometer with a geometrical acceptance covering 93\,\% of the full solid angle. The cylindrical core of the detector consists of four main components. A helium-based (60\,\% He, 40\,\% C$_3$H$_8$) multi-layer drift chamber (MDC) provides a charged-particle momentum resolution of 0.5\,\% at 1\,GeV/$c$ in a 1\,T magnetic field as well as specific energy loss (d$E$/d$x$) measurements with a resolution better than 6\,\% for electrons from Bhabha scattering. Particle identification  is provided by a plastic scintillator time-of-flight system (TOF) with a time resolution of 68\,ps in the barrel region. The time resolution of the end cap TOF system was 110\,ps for the data taken before 2015, which was after the upgrade with multi-gap resistive plate chamber technology improved to 60\,ps in 2015~\cite{ToF,ToF_upgrade}. Photons are measured using a CsI (Tl) electromagnetic calorimeter (EMC) with an energy resolution of 2.5\,\% (5\,\%) at 1\,GeV in the barrel (endcap) region. The 1.0\,T magnetic field is provided by a superconducting solenoid magnet. It is supported by an octagonal flux-return yoke with resistive plate counter muon identifier modules interleaved with steel, by which muon tracks of momenta larger than 0.5\,GeV/$c$ are detected with a position resolution better than 2\,cm. More details on the BESIII detector can be found in Ref.~\cite{bes3}.
\begin{table}[h]
\begin{center}
  \caption{\label{tab:BeamEnergies} Summary of integrated luminosities $\cal{L}$~\cite{bes3_lumi_paper} of data sets at six center-of-mass 
energies~\cite{bes3_ecms_paper} analyzed.}
\begin{tabular}{c|c}
 $\sqrt{s}$\,[MeV]    & Luminosity\,[pb$^{-1}$]   \\\hline
  4178.0   &  3194.5      \\
  4226.3   &  1091.7    \\
  4258.0   &  ~825.7    \\
  4358.3   &  ~539.8    \\
  4415.6   &  1073.6    \\
  4599.5   &  ~566.9    \\
\end{tabular}
\end{center}
\vspace{-0.45cm}
\end{table}
\begin{table}[h]
\begin{center}
  \caption{\label{tab:DecayChans} Summary of the $16$ hadronic $\eta_{\rm c}$ decay channels under consideration.}
\begin{tabular}{c|c|c}
Decay                     & ${\cal B}_{i}$[\%]~\cite{PDG2016}        & Mode No. $i$ \\\hline
 $3(\pi^+\pi^-)$           &  1.8  $\pm$ 0.4   &  01 	\\
 $2(\pi^+\pi^-\pi^0)$      & 17.4 $\pm$ 3.3    & 02  	\\
 $\pi^+\pi^-\pi^0\pi^0$    &  4.7  $\pm$ 1.0   & 03  	\\
 $2(\pi^+\pi^-)$           & 0.97 $\pm$ 0.12   & 04  	\\
 $K^0_SK^+ \pi^-$            & 2.43 $\pm$ 0.17   & 05  	\\
 $K^+K^-\pi^+\pi^-$        & 0.69 $\pm$ 0.11   & 06  	\\
 $K^+K^-\pi^0$             & 1.21 $\pm$ 0.83   & 07  	\\
 $K^0_SK^+\pi^-\pi^+\pi^-$   & 2.75 $\pm$ 0.74   & 08  	\\ 
 $2(\pi^+\pi^-)\eta$	   &  4.4 $\pm$ 1.3    & 09     \\
 $\pi^+\pi^-\eta$	   &  1.7 $\pm$ 0.5    & 10     \\
 $K^+K^-\eta$  	           &  1.35 $\pm$ 0.16  & 11     \\
 $K^+K^-K^+K^-$	           & 0.146 $\pm$ 0.030 & 12     \\
 $K^+K^-2(\pi^+\pi^-)$     & 0.75 $\pm$ 0.24   & 13     \\
 $p\bar{p}$		   & 0.150 $\pm$ 0.016 & 14     \\
 $p\bar{p}\pi^+\pi^-$      & 0.53 $\pm$ 0.18   & 15     \\
 $p\bar{p}\pi^0$	   & 0.36 $\pm$ 0.13   & 16     \\\hline
 \rule{0pt}{12pt}   Summed up  & $\sum_{i}\mathcal{B}_i= 41.34 \pm 3.93$    &  
\end{tabular}
\end{center}
\vspace{-0.45cm}
\end{table}

The analysis is based on  $e^+e^-$ annihilation data samples corresponding to an integrated luminosity of 7.3\,fb$^{-1}$, collected with the BESIII detector at six different center-of-mass energies $\sqrt{s}$ between 4.18\,GeV and 4.60\,GeV as listed in Tab.\,\ref{tab:BeamEnergies}. At each $\sqrt{s}$, we measure the cross sections of the three different exclusive reactions $e^+e^-\rightarrow \eta_{\rm c}\pi^+ \pi^-\pi^0$, $e^+e^- \rightarrow \eta_{\rm c}\pi^+ \pi^-$ and $e^+e^- \rightarrow \eta_{\rm c}\pi^0\gamma$, respectively. In the case of the observation of significant $\eta_{\rm c}$ production, we perform a search for a possible intermediate $Z_{\rm c}$ state decaying to $\eta_{\rm c} \pi$. We reconstruct in total 16 hadronic $\eta_{\rm c}$ decay channels as summarized in Tab.\,\ref{tab:DecayChans}, corresponding to about 40\% of the total $\eta_{\rm c}$ branching fraction. 

In order to determine the detection efficiencies and to study background contributions, a Monte Carlo (MC) simulation software based on {\textsc GEANT4}~\cite{Geant} is used that includes the geometrical BESIII detector description and response. The event selection criteria and the detection efficiencies are determined and studied based on samples of $1 \times 10^5$ signal events generated at each value of $\sqrt{s}$.  

In the simulation of the various exclusive event samples, the beam energy spread and the initial state radiation (ISR) in the $e^+e^-$ annihilation are included employing the {\textsc KKMC} generator~\cite{KKMC}. The inclusive MC simulated event samples comprise production of open-charm processes, ISR and hadronic production of light hadron and vector charmonium (-like) states as well as continuum processes, correspondingly for a given $\sqrt{s}$. The signal decay modes are modeled with {\textsc EVTGEN}~\cite{EvtGen}. Final state radiation (FSR) from charged final-state particles are incorporated by the {\textsc PHOTOS} package~\cite{PHOTOS}. 

\section{\label{sec:Ana} Data analysis}
The analyses of the three different exclusive $\eta_{\rm c}$-production channels are very similar, and the  $Z_{\rm c} \to \eta_{\rm c}\pi$ search merely differs by the additional corresponding mass window cut on the $\eta_{\rm c}$. The event selections (Sec.\,\ref{subsec:EvtSel}) and the determination of the reconstruction efficiencies (Sec.\,\ref{subsec:RecoEffi}) are essentially the same. The radiative corrections (Sec.\,\ref{subsec:RadCorr}) applied are slightly differently calculated, depending on whether or not a significant production cross section is measured, and thus the measured line shape can be used in an iterative procedure, or instead, an assumed line shape needs to be used.  
    
\subsection{\label{subsec:EvtSel} Event selection}
Several event selection criteria have been studied and applied. Charged tracks are reconstructed from the hits in the MDC within the polar-angle range of $|\cos\theta| < 0.93$. The tracks are required to have the point of closest approach to the interaction point within $\pm$10\,cm in the beam direction and within 1\,cm in the plane perpendicular to it. For each charged track, the TOF and the d$E$/d$x$ information are combined to calculate particle identification (PID) probabilities (based on $\chi^2_{\rm PID} = \chi^2_{\rm TOF} + \chi^2_{{\rm d}E/{\rm d}x}$) for $\pi$, $K$ and $p$ hypotheses. We assign that particle type to a track for which the largest probability is obtained. For $K_S^0$ candidates, all possible combinations of two oppositely charged tracks selected using the standard \mbox{BESIII} criteria~\cite{hcpipi_BESIII} and assumed to be pions are formed. Next, primary and secondary vertex fits~\cite{VtxKinFits} are performed and the decay length from the secondary vertex fit is required to be greater than twice the uncertainty. A 15\,MeV/$c^2$ mass window cut, corresponding to about $\pm 3\sigma$ around the nominal $K_S^0$ mass~\cite{PDG2016}, is applied. 

Photon candidates are required to have an energy deposit of at least 25\,MeV in the barrel (polar-angle region of $|\cos(\theta)| < 0.80$ with respect to the beam axis) and 50\,MeV in the endcap ($0.86 < |\cos(\theta)| < 0.92)$) region of the EMC. Timing requirements for EMC clusters are used to suppress electronic noise and energy deposits unrelated to the event. Moreover, the candidates are required to be at least $20^\circ$ away from the nearest charged track to reject EMC hits caused by split-off clusters from charged tracks. Decays of $\pi^0$ and $\eta$ to $\gamma\gamma$ are reconstructed from photon pairs and selected by mass window cuts also corresponding to about $\pm 3\sigma$ ($31$\,MeV and $67$\,MeV, respectively) around the nominal $\pi^0$ and $\eta$ mass~\cite{PDG2016}, to which also a one constraint (1C) kinematic fit is imposed to improve resolutions.

A kinematic 4C fit is performed, constraining the four momenta to $\sqrt{s}$  (and, where applicable, imposing one additional constraint for each $\pi^0$ and $\eta$ in the decay). In the case of multiple final-state  candidates, that one with the minimum $\chi^2 = \chi^2_{\rm 4C} + \chi^2_{\rm PID}$ is selected. Both the vertex and the kinematic fit are required to satsify  $\chi^2 < 100$.

Since at the collision energies under consideration it is not possible for both a charmed meson and an $\eta_{\rm c}$ to be produced, events are rejected if a $D$-meson candidate is reconstructed in one of the five major decay modes (including charge conjugates): $D^0\to K^-\pi^+$, $D^0\to K^-\pi^+\pi^0$, $D^+\to K^- \pi^+ \pi^+$, $D^+\to K^0_S\pi^+$ and $D^+ \to K^0_S\pi^+\pi^0$. Similarly, events are rejected with a $K^*(892) \to K\pi$, an $\omega \to \pi^+\pi^-\pi^0$ or an $\eta \to \pi^+\pi^-\pi^0$ candidate in the final state. These veto cuts are optimized using the figure of merit ${\rm FoM} = S/\sqrt{B}$, corresponding to a maximized significance in case of small signals. The signal $S$ is the number of events obtained from the signal MC simulation and the background $B$ is estimated based on the sidebands outside the $\eta_{\rm c}$ mass window in the data. If the gain $\Delta {\rm FoM_{\rm veto}}$ by an optimal veto cut is at least $\Delta {\rm FoM_{\rm veto}}/{\rm FoM_0} = 1.5\,\%$ with respect to the ${\rm FoM_0}$ without a veto cut, the cut is applied. The veto cut mass ranges applied correspond to about $2\sigma$ for all cases, and for $e^+e^-\rightarrow \eta_{\rm c}\pi^+ \pi^-\pi^0$ for example, at least one veto cut is applied to each of the 16 hadronic decay channels; in total up to 38 veto cuts are applied here. For the $Z_{\rm c}$ search, an additional selection is applied in terms of an invariant mass window cut for the underlying $\eta_{\rm c}$ of $2.880\,$GeV/$c^2< m_{\rm \eta_{\rm c,cand.}}<3.080$\,GeV/$c^2$. 

The $\eta_{\rm c}$ and $Z_{\rm c}$ invariant-mass spectra for each individual exclusive final state are constructed by adding up the mass spectra of all possible combinations within this final state for a given $\eta_{\rm c}$ decay channel, forming an $\eta_{\rm c}$ or a $Z_{\rm c}$ candidate. In the reaction of $e^+e^- \to \eta_{\rm c}\pi^+\pi^-\pi^0$ and the $\eta_{\rm c} \to 2(\pi^+\pi^-)$ decay channel for example, the mass spectra of the nine different $2(\pi^+\pi^-)$ combinations within the $3(\pi^+\pi^-)\pi^0$ final state are summed up.
%
%
\begin{figure*}[tp!]
    \begin{center}
     \includegraphics[clip,trim= 5 0 12 5, width=0.32\linewidth, angle=0]{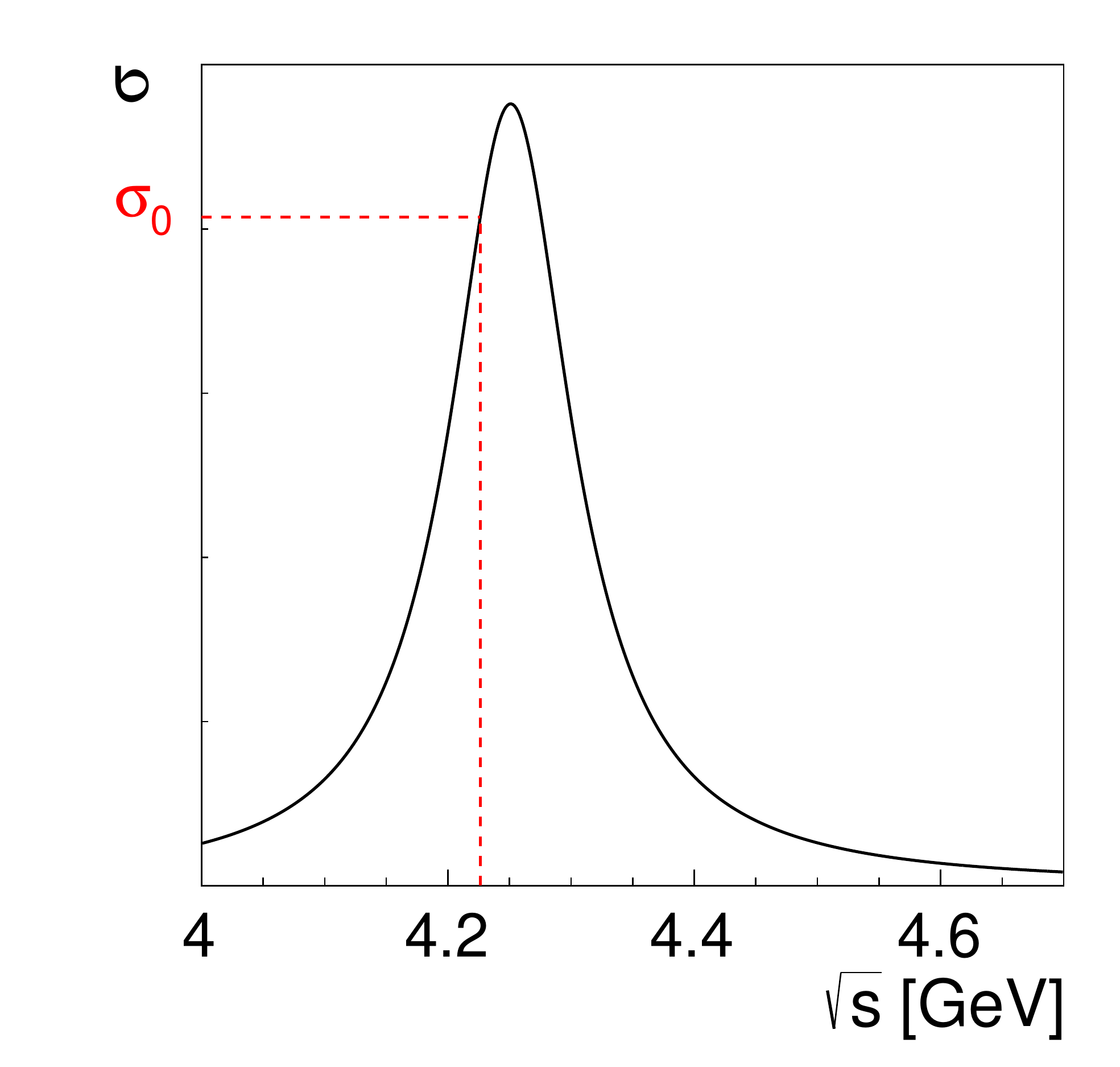}
     \includegraphics[clip,trim= 5 0 12 5, width=0.32\linewidth, angle=0]{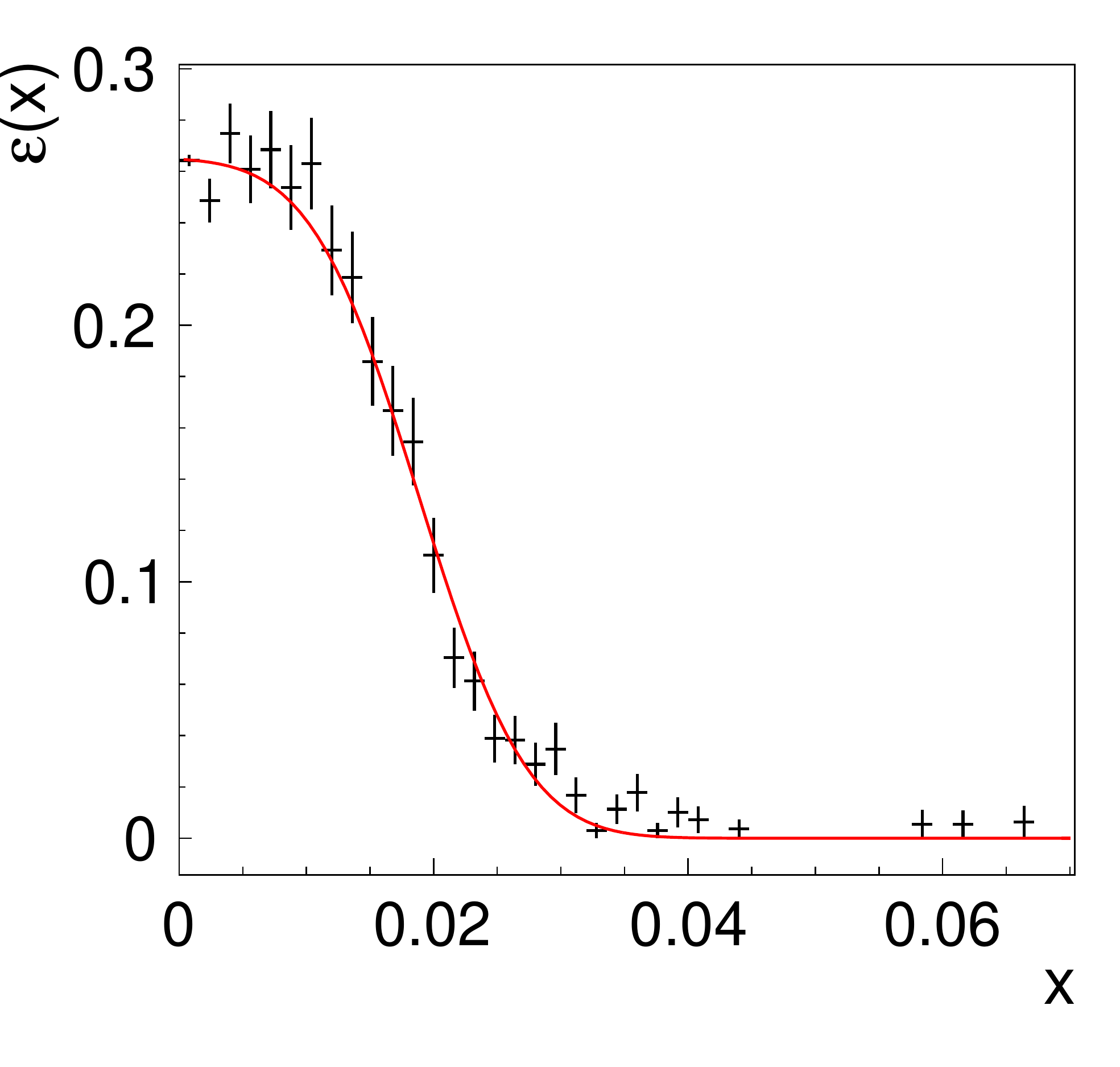}
     \includegraphics[clip,trim= 5 0 12 5, width=0.32\linewidth, angle=0]{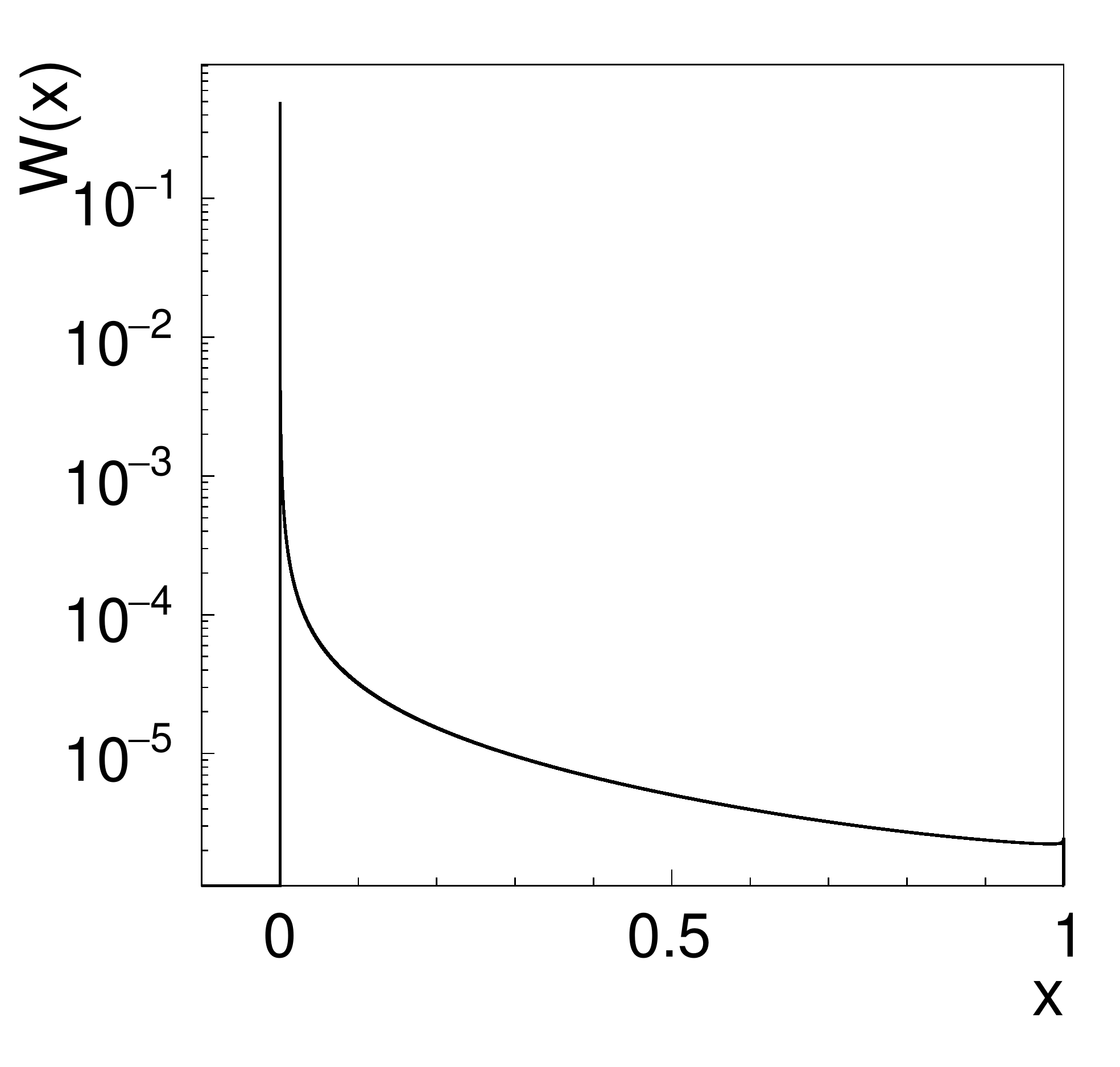}
     \caption{Radiative corrections: line shape of an assumed underlying resonance, here the $Y(4260)$ with parameters 
       $(m, \Gamma_0) = (4.251\,{\rm GeV}/c^2, 0.120\,{\rm GeV})\,$~\cite{PDG2016} with the cross section $\sigma_0$ 
       indicated at $\sqrt{s}=4.23$\,GeV {\it (left)}; a reconstruction-efficiency distribution extracted 
       from signal MC simulations as a function of the ISR photon energy fraction $x=E_\gamma/E_{\rm beam}=2\cdot E_\gamma/\sqrt{s}$, 
       shown for the example of $e^+e^- \to \eta_{\rm c}\pi^+ \pi^-\pi^0$, with $\eta_{\rm c} \rightarrow p\bar{p}$, at $\sqrt{s}=4.23$\,GeV 
       {\it (center)}; and the radiator function $W(x)$ {\it (right)}.}
      \label{fig:RadCorrExplain}
     \end{center}
\end{figure*}   
  
\subsection{\label{subsec:RecoEffi} Reconstruction efficiencies}
The line shapes of the $\eta_{\rm c}$ ($Z_{\rm c}$) signals in the different $\eta_{\rm c}$ decay channels are determined based on the signal MC simulated data. Using a fit of a Voigtian (a convolution of a Breit-Wigner with a Gaussian) to the reconstructed truth matched mass distributions, in which the width parameter $\Gamma$ is fixed to the nominal PDG value (and assumed Breit-Wigner parameters for the $Z_{\rm c}$), we determine the resolution parameters as well as possible mass shifts individually and take them into account. The extracted parameters are fixed in both the fits to the signal MC simulations, used to determine the reconstruction efficiencies, and in the simultaneous fits to the data (Sec.\,\ref{sec:Xsec}). The reconstruction efficiencies are determined by the ratio of the reconstucted events (integral of the signal part of the fit function) over the number of generated events.
\subsection{\label{subsec:RadCorr} Radiative corrections}
Initial state radiation (ISR) leads to energy losses of the initial $e^+ e^-$ system via emission of bremsstrahlung. As a consequence, at each nominal energy point the data are produced over a range of center-of-mass energies rather than at a fixed value, and the reconstruction efficiencies thus depend on the fraction $x=E_\gamma/E_{\rm beam}$ of the emitted photon energy with respect to the beam energy. In order to provide measured Born cross sections $\sigma_{\rm Born}$ in addition to the experimentally observed cross sections $\sigma$ (Sec.\,\ref{subsec:etacpipipi0}) and also to provide most conservative 90\,\% C.L. upper limits UL$_{\rm 90}$ (Sec.\,\ref{subsec:etacULS}, Sec.\,\ref{sec:Zc}), ISR corrections are calculated as follows.
%
%
\begin{figure*}[tp!]
    \begin{center}
     \includegraphics[clip,trim= 0 0 0 10, width=0.325\linewidth, angle=0]{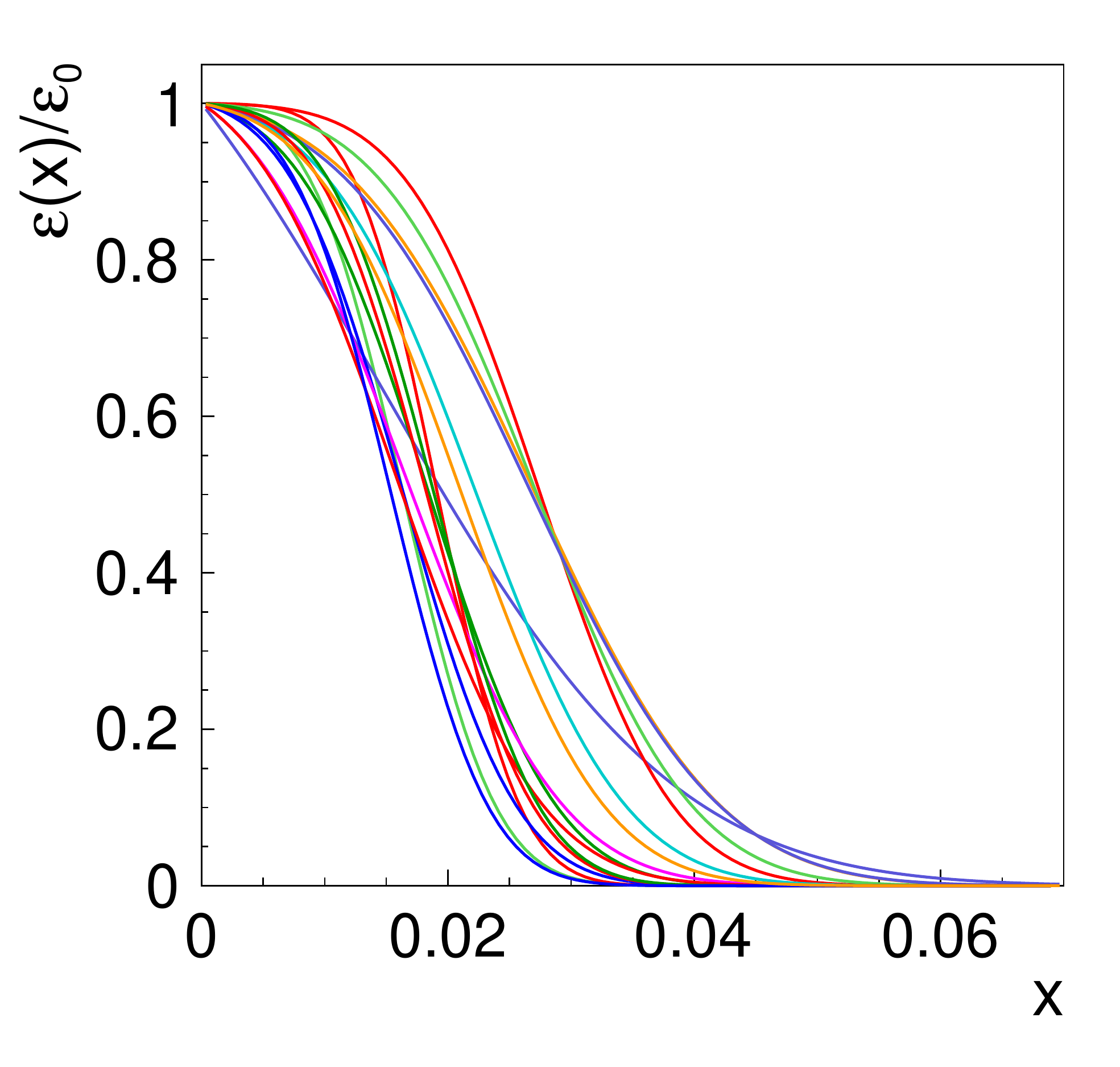}
     \includegraphics[clip,trim= 0 0 0 10, width=0.325\linewidth, angle=0]{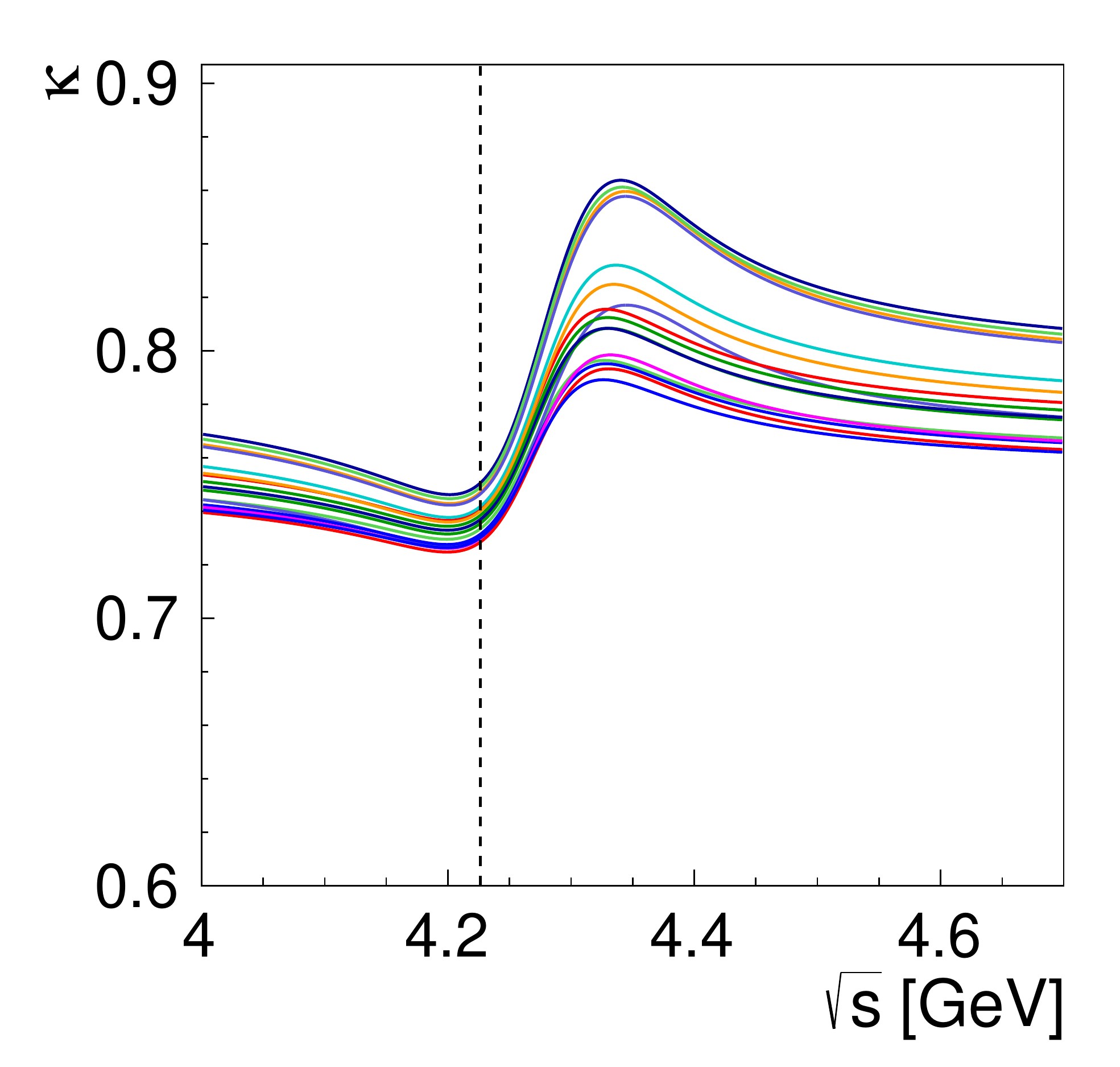}
     \includegraphics[clip,trim= 0 0 0 10, width=0.325\linewidth, angle=0]{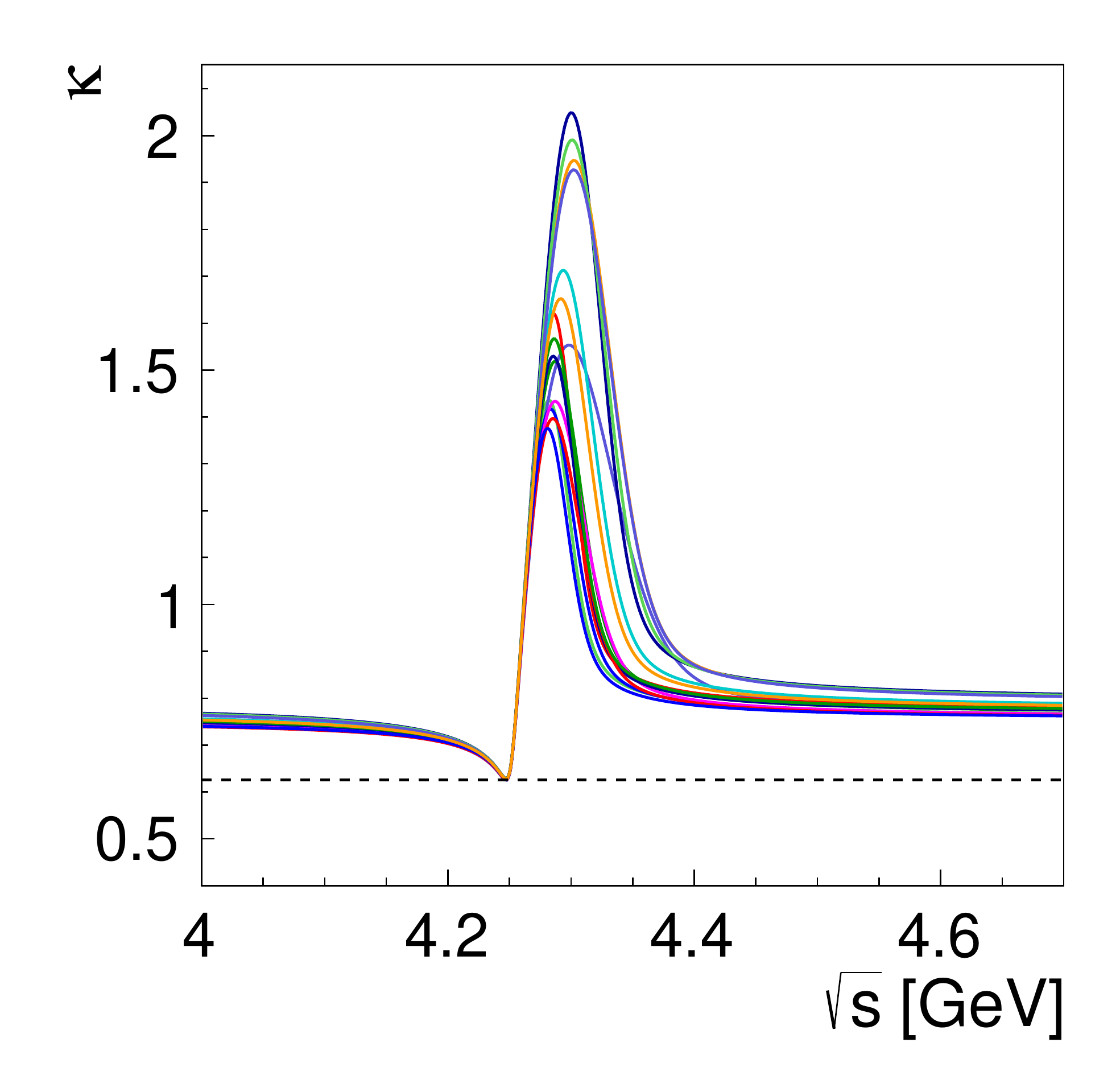}
     \caption{Normalized efficiency distributions $\epsilon(x)/\epsilon_0$ as a function of $x=E_\gamma/E_{\rm beam}$ shown 
       for the $16$ $\eta_{\rm c}$ decay channels of the example of $e^+e^- \to \eta_{\rm c}\pi^+ \pi^-\pi^0$ at 
       $\sqrt{s}=4.23$\,GeV {\it (left)}. Corresponding radiative corrections $\kappa_{i}$ as a 
       function of $\sqrt{s}$ for an intermediate resonance with $m=4.251$\,GeV/$c^2$ for two different 
       assumed natural widths of $\Gamma_{\rm 0,PDG} = 120$\,MeV {\it (center)} and $\Gamma_{\rm 0,UL}=10$\,MeV 
       {\it (right)}, respectively.
}
     \label{fig:RadCorrExplain_b}
   \end{center}
\end{figure*}
 
The photon-energy fractions $x$ are distributed according to the radiator function $W(x)$ as defined in Ref.~\cite{radFctn}. The total number of observed events for a decay channel with branching ratio ${\mathcal B}$ is given by 
\begin{equation}
        N = \mathcal{L} \cdot {\mathcal B} \cdot \int_{0}^{1}{\sigma(x) \epsilon(x) W(x) {\rm d}x}~,
\label{Eq.radCorr_a}
\end{equation}
where $\mathcal{L}$ is the integrated luminosity, $\sigma(x)$ is the production cross section, and $\epsilon(x)$ the reconstruction efficiency. The cross section or line shape $\sigma(x)$ could be the one of an underlying resonance, such as the $Y(4260)$~\cite{PDG2016} (Fig.\,\ref{fig:RadCorrExplain}, {\it left}) that is assumed to be produced in direct formation, and that subsequently decays to the considered production channel of $\eta_{\rm c}$ (or $Z_{\rm c}$) plus corresponding recoil particles. 

As an example, one of the $16$ reconstruction efficiency curves $\epsilon_{i}(x)$ for the case of $e^+e^- \to \eta_{\rm c}\pi^+ \pi^-\pi^0$, with $\eta_{\rm c} \rightarrow p\bar{p}$ (which according to Tab.\,\ref{tab:DecayChans} corresponds to mode number $i$ = 14), at $\sqrt{s}=4.23$\,GeV is shown (Fig.\,\ref{fig:RadCorrExplain}, {\it center}) as well as the radiator function $W(x)$ (Fig.\,\ref{fig:RadCorrExplain}, {\it right}). The Born cross section $\sigma_{\rm Born}$ and the efficiency $\epsilon_{\rm 0}=\epsilon(x$=$0)$ at the nominal center-of-mass energy can be factored out. One rewrites Eq.\,(\ref{Eq.radCorr_a}) and introduces the radiative correction factor $\kappa$ defined as
\begin{eqnarray}
        N &=& \mathcal{L} \cdot {\mathcal B} \cdot \sigma_{\rm Born} \cdot \epsilon_{\rm 0} \cdot \int_{0}^{1}{{\frac{\sigma(x)}{\sigma_{\rm Born}} \frac{\epsilon(x)}{\epsilon_{\rm 0}}  W(x)} {\rm d}x}~ \nonumber\\
   \kappa &:=& \int_{0}^{1}{{\frac{\sigma(x)}{\sigma_{\rm Born}} \frac{\epsilon(x)}{\epsilon_{\rm 0}}  W(x)} {\rm d}x}~,\nonumber
\label{Eq.radCorr_kappa}
\end{eqnarray}
so that Eq.\,(\ref{Eq.radCorr_a}) can be formulated in terms of the Born cross section $\sigma_{\rm Born}$:
\begin{equation}
        N = \mathcal{L} \cdot {\mathcal B} \cdot \sigma_{\rm Born} \cdot \epsilon_{\rm 0} \cdot \kappa~.
\label{Eq.radCorr_b}
\end{equation}

The $\epsilon(x)$ distributions and \mbox{$\epsilon_{\rm 0}$}, {\it i.e.} the reconstruction efficiency for the case of no ISR effect, are obtained using signal MC simulations with and without the ISR effect being enabled in the {\textsc KKMC} generator. As illustrated in Fig.\,\ref{fig:RadCorrExplain_b}, where the $\epsilon_i(x)/\epsilon_{i,0}$ distributions corresponding to the 16 different $\eta_{\rm c}$ decay channels are shown for the example of $\eta_{\rm c}\pi^+ \pi^- \pi^0$ at $\sqrt{s}=4.23$\,GeV (Fig.\,\ref{fig:RadCorrExplain_b}, {\it left}), the correction factors $\kappa_i$ depend on $\sqrt{s}$. They are shown for the same $\eta_{\rm c}$ example for two different assumed widths of an underlying resonance (Fig.\,\ref{fig:RadCorrExplain_b}, {\it center/right}). The shape of $\kappa_i(\sqrt{s})$ significantly depends on the line shape of the resonance, in particular the minimum of the $\kappa_i$ distributions is lower for more narrow resonances.

When the energy-dependent cross section is roughly known or can be measured, as in the case of $\sigma_{\rm B}(e^+e^- \to \eta_{\rm c}\pi^+\pi^-\pi^0)$ (Sec.\,\ref{subsec:etacpipipi0}), we determine the $\kappa_i(\sqrt{s})$ for each $\sqrt{s}$ by applying an iterative procedure. Initially, a default input cross section $\sigma_0(\sqrt{s})$, namely the one of the $Y(4260)$ modeled as a Breit-Wigner shaped resonance with parameters taken from Ref.~\cite{PDG2016} (Fig.\,\ref{fig:RadCorrExplain}, {\it left}), is used to measure the cross section at each individual $\sqrt{s}$. This is done using the corresponding normalized efficiency curves (Fig.\,\ref{fig:RadCorrExplain_b}, {\it left}), and applying the initially calculated numbers $\kappa_{i}(\sqrt{s})$ (Fig.\,\ref{fig:RadCorrExplain_b}, {\it center}). A fit to the resultant cross section distribution (after iteratively applied radiative corrections) is then used as an updated input $\sigma'(\sqrt{s})$, and the procedure is repeated until the line shape converges.

In the case where no significant cross section is observed,  we follow the most conservative approach to provide upper limits (Sec.\,\ref{subsec:etacULS}, Sec.\,\ref{sec:Zc}) and assume for all six $\sqrt{s}$ points the minimum correction factors $\kappa_{i, {\rm min}}$ based on an artificially narrow natural input width $\Gamma_{\rm 0,UL}=10\,$MeV for the resonance (Fig.\,\ref{fig:RadCorrExplain_b}, {\it right}). In the example shown (Fig.\,\ref{fig:RadCorrExplain_b}, {\it right}) the $\kappa_{i, {\rm min}}$ values are about 0.6 for all decay channels.

%
%
\section{\label{sec:Xsec} cross section measurements for $\eta_{\rm c}$ production}
In order to measure the $\eta_{\rm c}$ and $Z_{\rm c}$ production cross sections and upper limits, we perform for each recoil system ($\pi^+\pi^-\pi^0$, $\pi^+\pi^-$ or $\pi^0\gamma$) and center-of-mass energy a maximum-likelihood fit simultaneously for all $\eta_{\rm c}$ decay channels to determine the common cross section  $\sigma$ from the data sets. The signal is described by a signal shape which is extracted from the signal MC simulation (Sec.\,\ref{subsec:RecoEffi}). The backgrounds are modelled by second to fourth-order polynomials. The fit function also allows for the presence of a $J/\psi$ signal in the data, which does not affect the main analysis results.
  
%
%
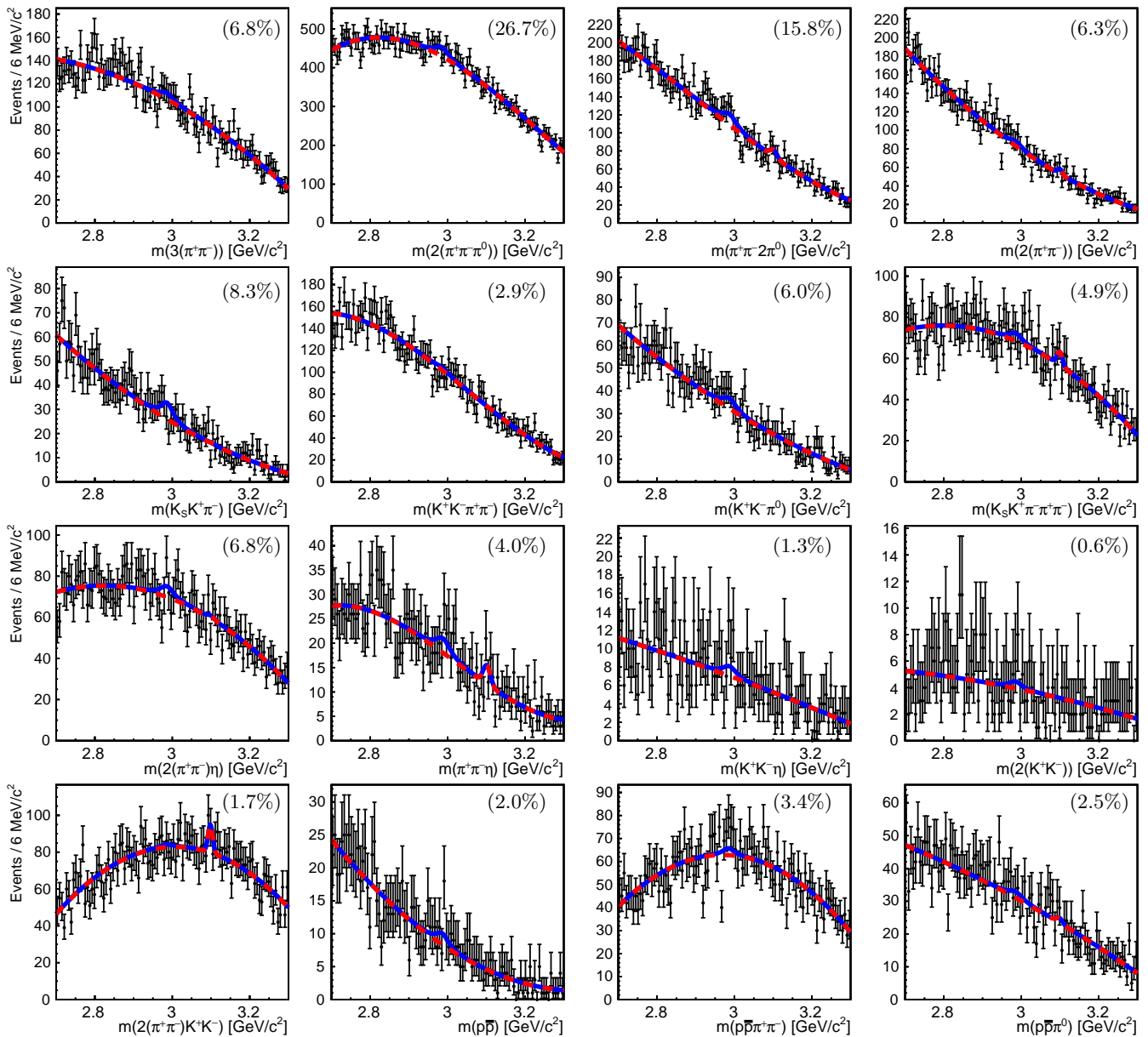
\begin{figure*}[tp!]
    \begin{center}
     \includegraphics[clip,trim= 0 0 0 0, width=1.0\linewidth, angle=0]{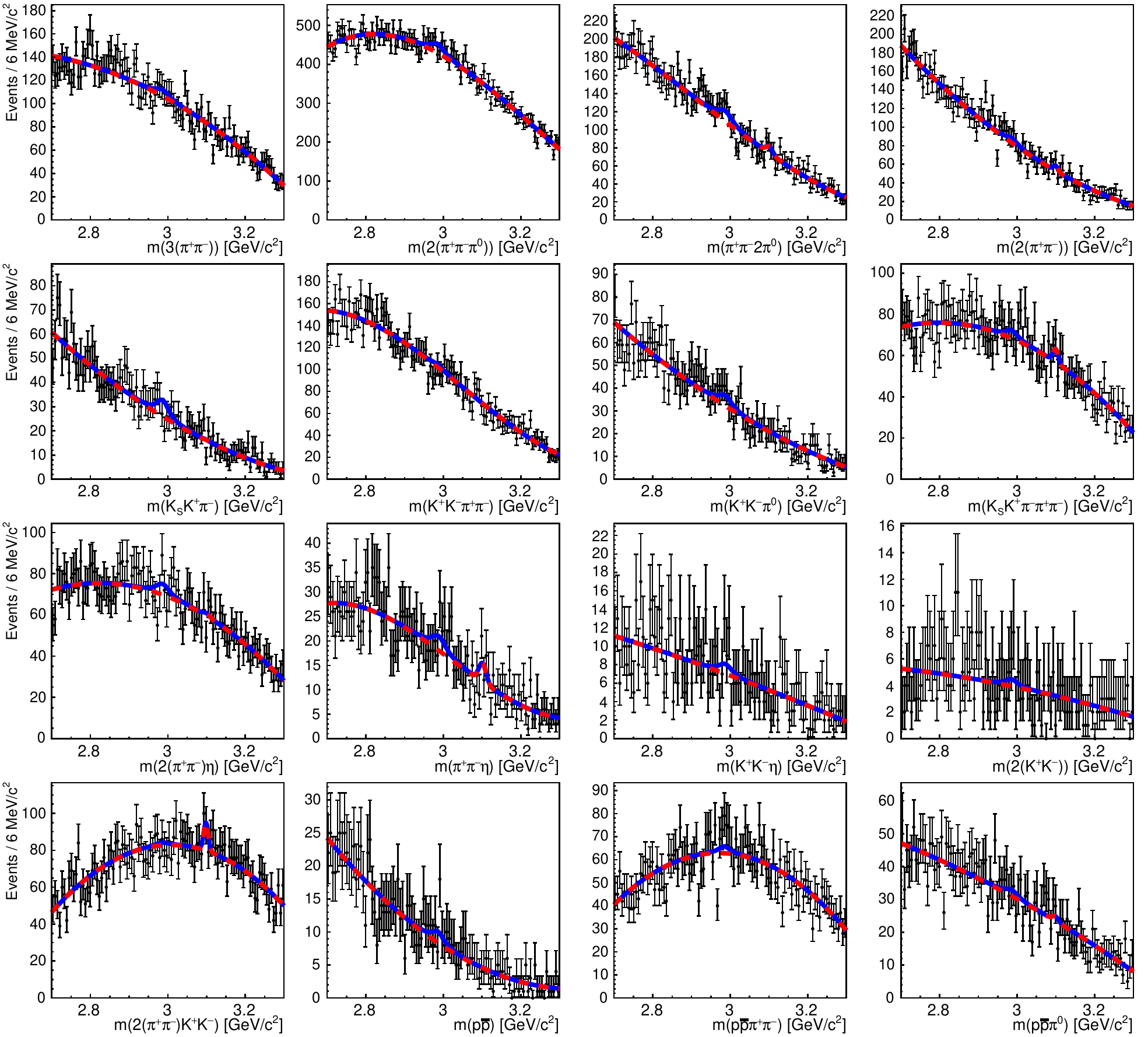}
     \caption{Simultaneous fit result for $\eta_{\rm c}$ production, with radiative corrections included, for the example case of $e^+e^- \to \eta_{\rm c}\pi^+\pi^-\pi^0$  at $\sqrt{s}=4.23\,$GeV. 
       The mass spectra are shown for the 16 $\eta_{\rm c}$ decay channels together with the simultaneous fit (blue solid curve). A moderate $\eta_{\rm c}$ peak is visible above the background (red dashed curve) for the majority of the spectra. The relative importance of each of the 16 decay channels in terms of $({\cal B}_{i} \times \epsilon_{i}) / \sum_{i=1}^{16}{ ({\cal B}_{i} \times \epsilon_{i}})$ is quoted for each corresponding mass spectrum.}
      \label{fig:simuFit_etac_M8_4230_bg3_ISR}
  \begin{picture}(1,1)
 \put(-158,511){(6.8\%)}
 \put(-40,511){(26.7\%)}
 \put( 87,511){(15.8\%)}
 \put(219,511){(6.3\%)}
 \put(-158,394){(8.3\%)}
 \put(-40,394){(2.9\%)}
 \put( 87,394){(6.0\%)}
 \put(219,394){(4.9\%)}
 \put(-158,281){(6.8\%)}
 \put(-40,281){(4.0\%)}
 \put( 87,281){(1.3\%)}
 \put(219,281){(0.6\%)}
 \put(-158,167){(1.7\%)}
 \put(-40,167){(2.0\%)}
 \put( 87,167){(3.4\%)}
 \put(219,167){(2.5\%)}
  \end{picture}
     \end{center}
\end{figure*}

Each of the fits minimizes the corresponding negative log-likelihood ($-\log({\rm LH})$). Accordingly, the branching ratios ${\cal B}_{i}$ and reconstruction efficiencies $\epsilon_{i}$ are taken into account, so that the  signal event yields in each channel, $N_{i}$, and the common cross section $\sigma$ are related as follows
\begin{equation}
        N_{i} = \sigma \cdot \mathcal{L} \cdot \epsilon_{i} \cdot {\mathcal B}_{i} ~~~\Leftrightarrow~~~~  
        \sigma  = \frac{N_{i}}{\mathcal{L} \cdot \epsilon_{i} \cdot {\mathcal B}_{i}}~.
\label{Eq.simuFit}
\end{equation}
Possible ISR energy losses are ignored here, so that the observed common cross section $\sigma$ is obtained, which is uncorrected for ISR effects and from which the number of observed events $N_{\rm obs}$ is obtained. 

In the case of the common dressed Born cross section $\sigma_{\rm Born}$, the corresponding radiative-correction factors $\kappa_{i}(\sqrt{s})$ (or $\kappa_{{\rm min},i}$) (Sec.\,\ref{subsec:RadCorr}) are included in the simultaneous fit, and the relation between $N_{i}$ and $\sigma_{\rm Born}$ then reads   
\begin{equation}
\sigma_{\rm Born} = \frac{N_{i}}{\mathcal{L} \cdot \epsilon_{{\rm 0},i} \cdot {\mathcal B}_{i}\cdot \kappa_{i}}~.
\label{Eq.simuFit_ISR}
\end{equation}
The $\epsilon_{i}$ in Eq.\,(\ref{Eq.simuFit}) are replaced by the $\epsilon_{{\rm 0},i}$ multiplied by the radiative correction factors $\kappa_{i}$ (or $\kappa_{{\rm min},i}$) (Sec.\,\ref{subsec:RadCorr}).  

Figure~\ref{fig:simuFit_etac_M8_4230_bg3_ISR} shows the $16$ reconstructed invariant mass spectra from the data  for the example channel  $e^+e^- \to \eta_{\rm c}\pi^+\pi^-\pi^0$ at $\sqrt{s} = 4.23$\,GeV,  together with the results of the simultaneous fit performed according to Eq.\,(\ref{Eq.simuFit_ISR}). A small $\eta_{\rm c}$ signal is visible in many of the spectra above the background function.
 
\begin{figure}[tp!]
    \begin{center}
     \includegraphics[clip,trim= 0 0 0 0, width=0.49\linewidth, angle=0]{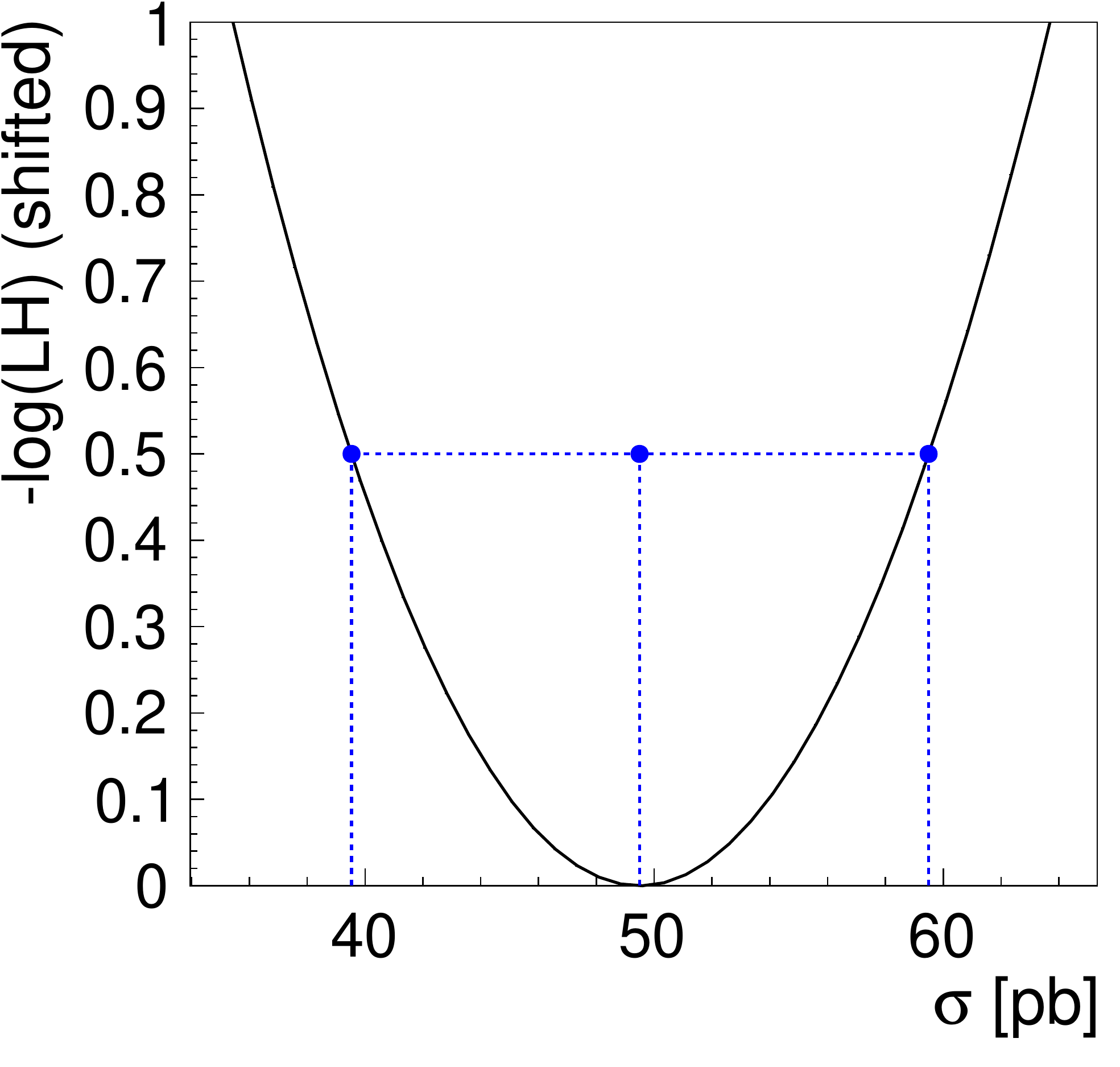}
     \includegraphics[clip,trim= 0 0 0 0, width=0.49\linewidth, angle=0]{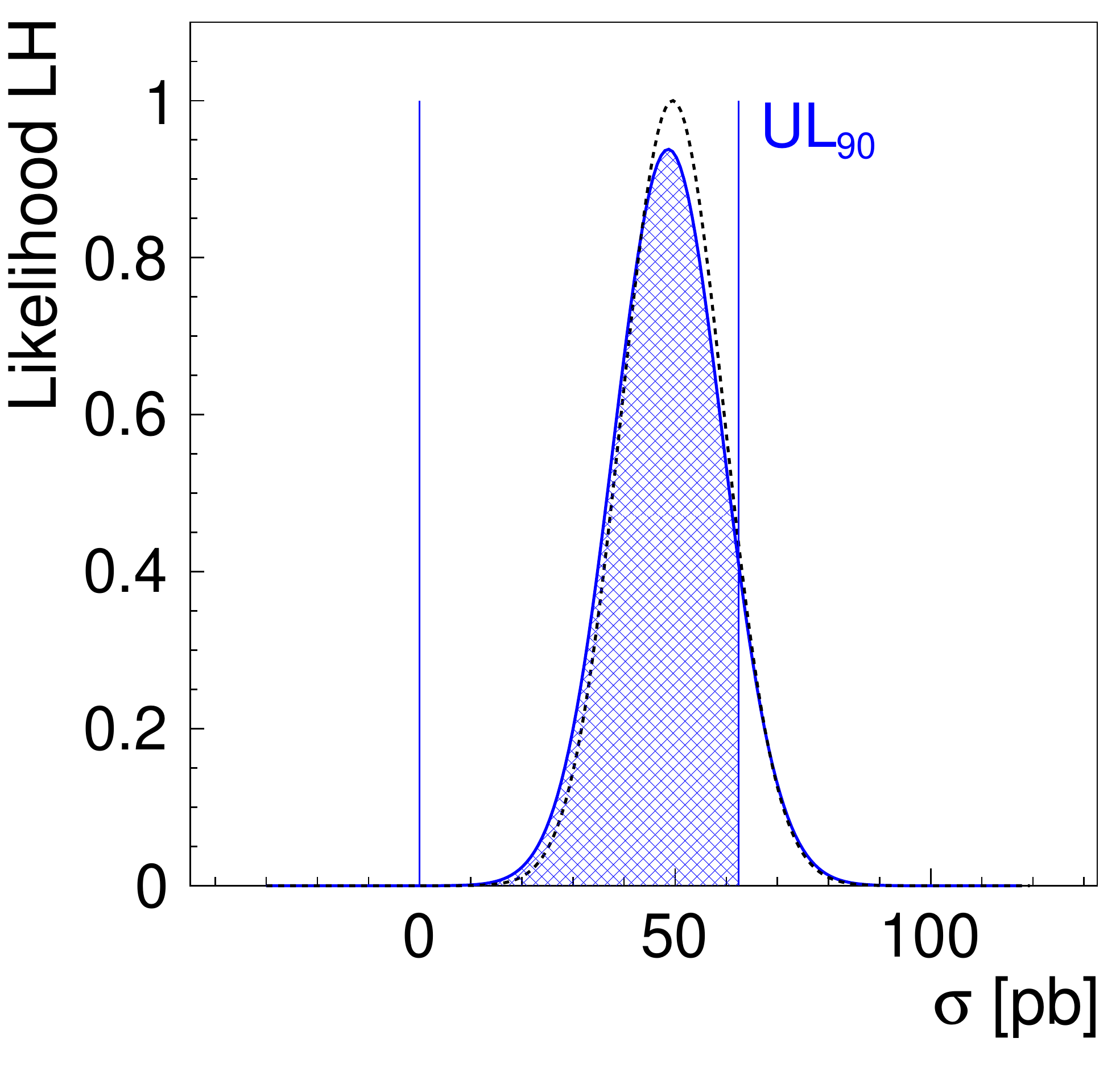}
     \vspace{-0.15cm}
     \caption{Example of a negative log-likelihood $-\log({\rm LH})$ {\it (left)} and the corresponding likelihood (LH) {\it (right)} 
       distributions, as obtained from the simultaneous fit, shown for the example of $e^+e^- \to \eta_{\rm c}\pi^+\pi^-\pi^0$ at 
       $\sqrt{s}=4.23\,$GeV, with radiative corrections included. 
       It is shown how the fit results with the corresponding statistical uncertainties are obtained from the negative log-likelihood, 
       and how the  90\,\% C.L. upper limits UL$_{\rm 90}$ are calculated from the likelihood, taking into account the systematic 
       uncertainty (blue, solid line).
}
      \label{fig:NegLogLikelihood_etac_M8_4230}
     \end{center}
\end{figure}

The  $-\log({\rm LH})$ curves allow the cross sections to be determined and the accompanying (potentially asymmetric)  statistical uncertainties,  as illustrated  in Fig.\,\ref{fig:NegLogLikelihood_etac_M8_4230}, ({\it left}). In the cases where no significant cross section is observed, 90\,\% C.L. upper limits (UL$_{\rm 90}$) are extracted  by integration of the corresponding convolved likelihood distributions, namely the area above zero and up to 90\% fraction (Fig.\,\ref{fig:NegLogLikelihood_etac_M8_4230}, {\it right}). For completeness, corresponding UL$_{\rm 90}$ values on the Born cross section are provided for all simultaneous fit results.
%
%

Vacuum polarization effects are accounted for by an energy-dependent correction factor $f_{\rm VP}$,  calculated according to Ref.~\cite{VacPol}, and applied to the dressed cross section values to obtain the final (undressed and bare) $\sigma_{\rm Born}$ values. The $f_{\rm VP}$ values are quoted for each $\sqrt{s}$ in the tables summarizing the $\eta_{\rm c}$ results (Tabs.\,\ref{tab:ResultsM8}, \ref{tab:ResultsM7}, \ref{tab:ResultsM9}).

\subsection{\label{subsec:etacpipipi0}Measurement of $\sigma_{\rm Born}(e^+e^- \to \eta_{\rm c}\pi^+\pi^-\pi^0)$}
\begin{figure}[bp!]
    \begin{center}
     \includegraphics[clip,trim= 0 0 0 0, width=0.49\linewidth, angle=0]{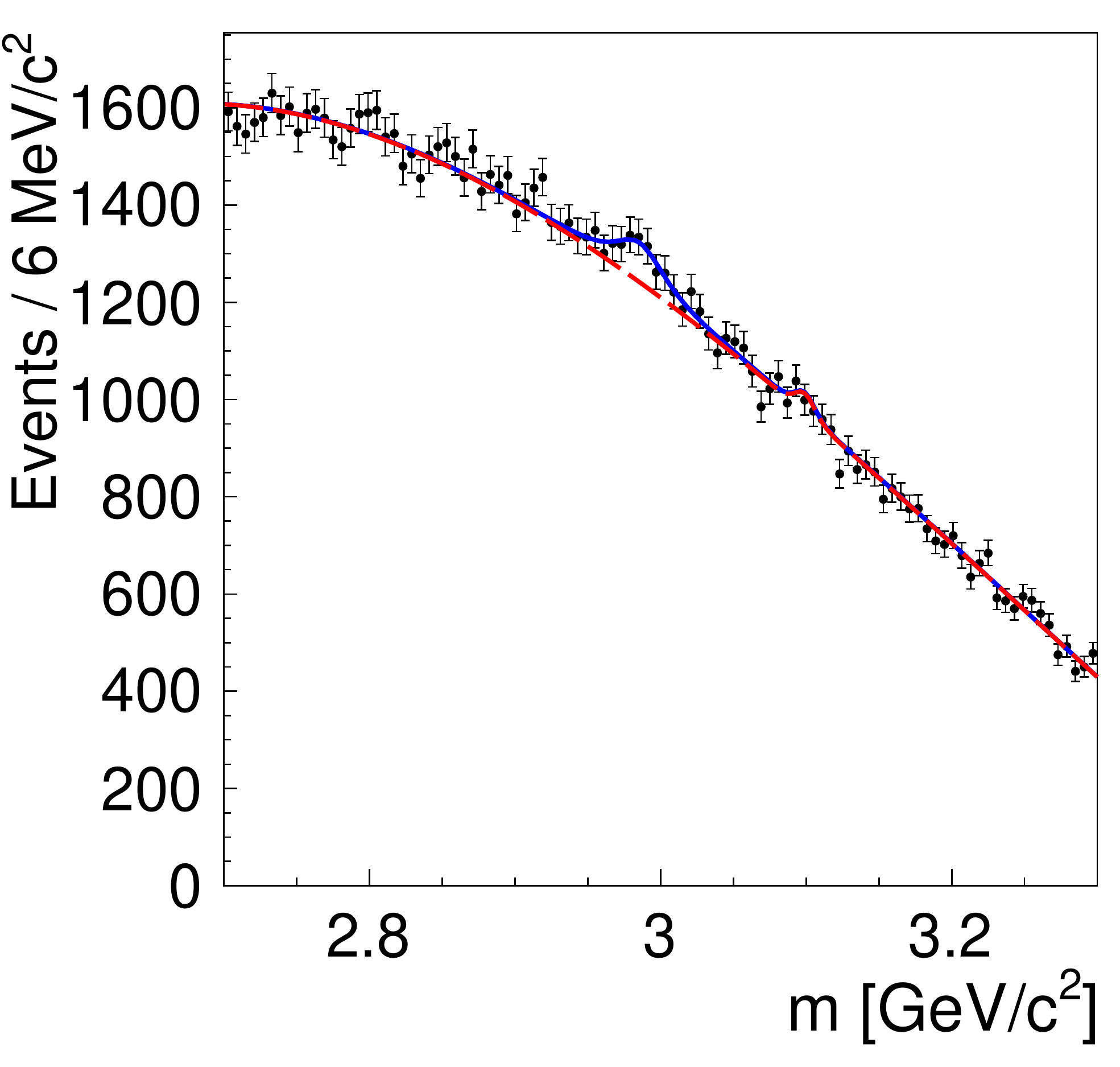}
     \includegraphics[clip,trim= 0 0 0 0, width=0.49\linewidth, angle=0]{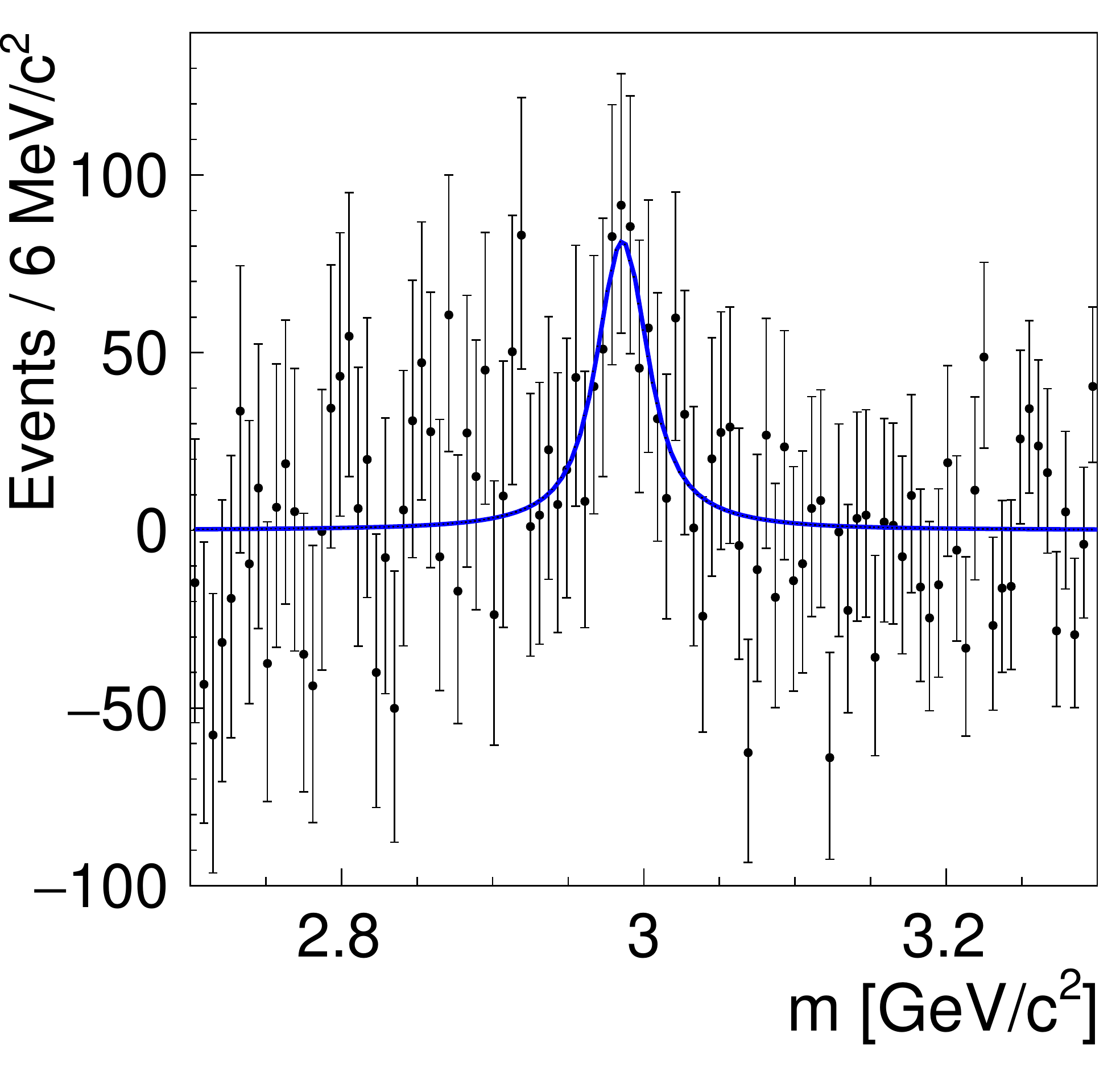}
     \vspace{-0.15cm}
     \caption{Reconstructed invariant mass distribution of the $\eta_{\rm c}$ candidates as summed from the 16 hadronic decay channels 
       analyzed in $e^+e^- \to \eta_{\rm c}\pi^+\pi^-\pi^0$ at $\sqrt{s}=4.23\,$GeV {\it (left)}, and the signal after background 
       subtraction {\it (right)}. The dots with error bars are the data, the solid (blue) lines are the total fit and the dashed 
       (red) line is the background description. A clear $\eta_{\rm c}$ peak is observed in the data.}
      \label{fig:simuFit_etac_M8_4230_proj}
     \end{center}
\end{figure}

The fit example shown in Fig.\,\ref{fig:simuFit_etac_M8_4230_bg3_ISR} is the result of the simultaneous fit of the common Born cross section (Tab.\,\ref{tab:DecayChans}) according to Eq.\,(\ref{Eq.simuFit_ISR}) for $\eta_{\rm c}\pi^+\pi^-\pi^0$ at $\sqrt{s}=4.23$\,GeV. Even though the signal is very small in each of the 16 $\eta_{\rm c}$ decay channels, it  becomes more distinct in the summed-up mass spectra, as seen in Fig.\,\ref{fig:simuFit_etac_M8_4230_proj}. This is in line with the measured cross section results obtained from the simultaneous fit combining the 16 $\eta_{\rm c}$ decay channels, as summarized in Tab.\,\ref{tab:ResultsM8}.

The statistical significances $\cal{S_{\rm stat}}$ are computed via the fraction of the integral of the likelihood distributions that lies below zero, namely the $p$-value defined as $\int_{-\infty}^{0}({\rm LH})/\int_{-\infty}^{\infty}({\rm LH})$. The significance is then expressed in terms of Gaussian standard deviations based on the inverse of the cumulative distribution function $\Phi$ of the standard normal distribution ${\cal N}(0,1)$, by computing ${\cal S}_{\rm stat} = \Phi^{-1}(1 -p)$.

We observe statistical significances of the reconstructed signals for this process that show a dependence on $\sqrt{s}$, ranging from about $2\sigma$ at $\sqrt{s}=4.18$\,GeV, increasing to a maximum of about $5\sigma$ at $\sqrt{s}=4.23$\,GeV and decreasing again down to less than $1\sigma$ at $\sqrt{s}=4.60$\,GeV. Since we measure a sizable cross section, the iterative procedure for computing the radiative correction factors $\kappa_i$ for each $\eta_{\rm c}$ decay channel is applied, as described in Sec.\,\ref{subsec:RadCorr}. The ranges of the finally applied $\kappa_i$ values are quoted as well as the number of observed events directly determined by the fit (Tab.\,\ref{tab:ResultsM8}).
%
%
%
%
%
\begin{figure}[tp!]
    \begin{center}
     \includegraphics[clip,trim= 0 0 0 0, width=0.65\linewidth, angle=0]{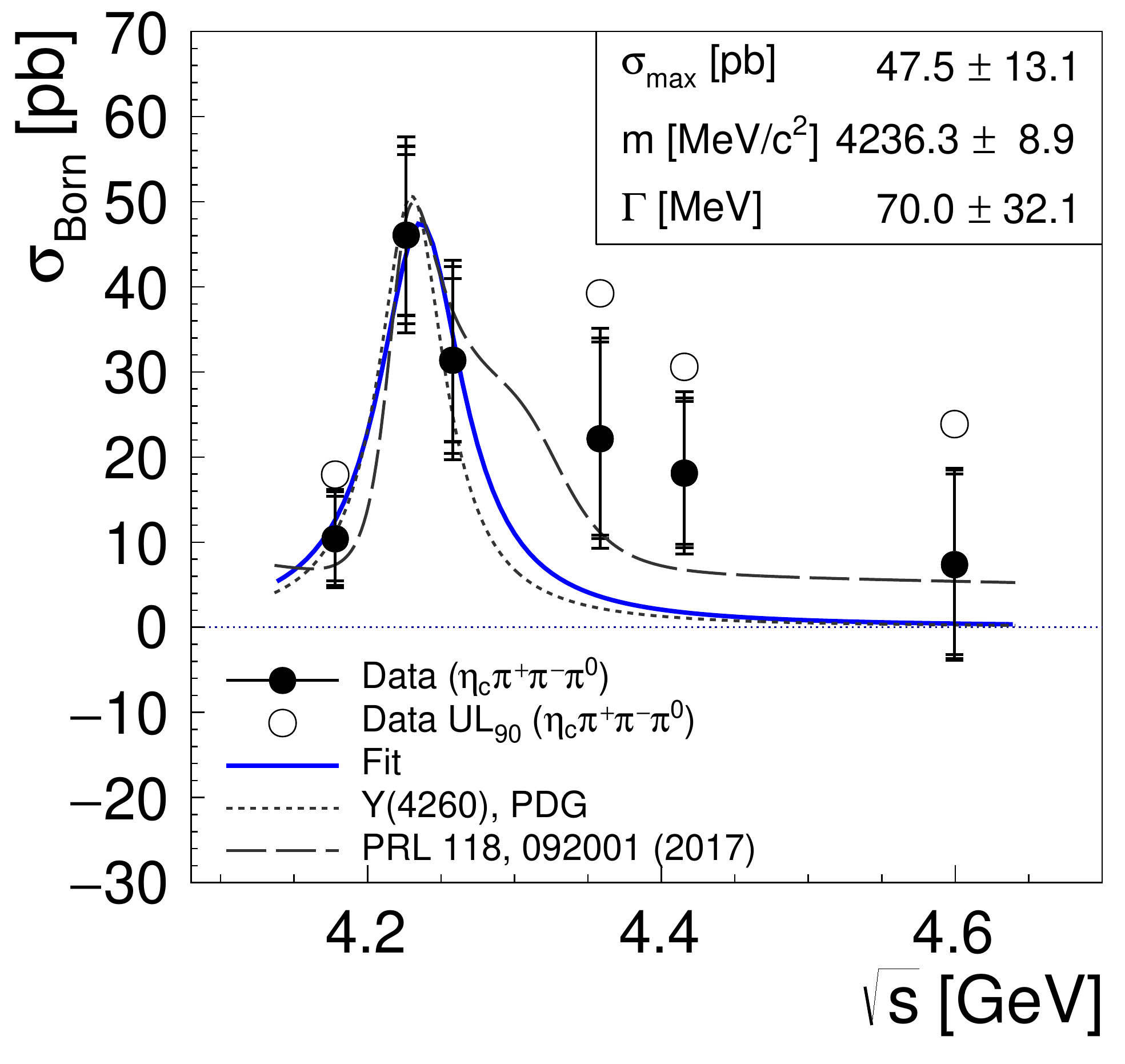}
     \caption{
       The measured cross sections for $e^+e^- \to \eta_{\rm c}\pi^+\pi^-\pi^0$ for different values of $\sqrt{s}$.
       The $\eta_{\rm c}$ production at $\sqrt{s}$ of 4.23\,GeV and 4.26\,GeV is observed with statistical significances 
       well above $3\sigma$. For the other center-of-mass energies, the 90\,\% C.L. upper limits are also shown (open circles).   
       Overlaid are a Breit-Wigner fit (blue curve) to the data points, the $Y(4260)$ line shape with the parameters found in the PDG~\cite{PDG18} 
       (dotted curve) and the lineshape as measured in the process 
       $J/\psi\pi^+\pi^-$ cross section~\cite{YshapeJpsipipi} (dashed curve). The three curves are found to be consistent with the 
       data, supporting the hypothesis of $\eta_{\rm c}\pi^+\pi^-\pi^0$ production via the $Y(4260)$.
     }
      \label{fig:EtacResults_etac_M8}
     \end{center}
\end{figure}

The results are graphically compiled in Fig.\,\ref{fig:EtacResults_etac_M8}. The data points at $\sqrt{s}$ of 4.23\,GeV and  4.26\,GeV each have a statistical significance of more than $3\sigma$. We fit a Breit-Wigner function to all six data points with the resultant resonance parameters $(m,\Gamma) = (4236.3\,{\rm MeV}/c^2 \pm 8.9\,{\rm MeV}/c^2, 70.0\,{\rm MeV} \pm 32.1\,{\rm MeV})$ that are compatible with those of the $Y(4260)$ resonance as given in the PDG~\cite{PDG18}. In addition to our fit, two $Y(4260)$  line shapes are overlaid, one with the PDG parameters, and one according to a recent measurement~\cite{YshapeJpsipipi}. The measured energy dependence of the  $e^+e^- \to \eta_{\rm c}\pi^+\pi^-\pi^0$ cross section is found to be consistent with that expected with the $Y(4260)$ participating as an intermediate resonance. In addition to the measured data points for all six center-of-mass energies, the corresponding upper limits are shown for those measured values of $\sigma_{\rm Born}$ that show a significance of less than $3\sigma$. 
\renewcommand{\arraystretch}{1.5}
\begin{table*}[bp!]
\begin{center}
  \caption{\label{tab:ResultsM8}
    Summary of results for $e^+e^- \rightarrow \eta_{\rm c} \pi^+\pi^-\pi^0$ based on the different $\sqrt{s}$ data sets~\cite{bes3_ecms_paper} 
    of integrated luminosities $\cal{L}$~\cite{bes3_lumi_paper}. Quoted are the number of observed events $N_{\rm obs}$ as obtained from 
    the simultaneous fits, the range of the applied radiative correction factors $\kappa_i$, the applied vacuum-polarization factor 
    $f_{\rm VP}$, the sum of the 16 efficiencies times branching ratio values $\sum\varepsilon_i{\cal B}_i$, the undressed Born cross 
    section $\sigma_{\rm Born}$ as well as the computed 90\,\% C.L. upper limits UL$_{\rm 90}$, and the computed statistical significance 
    including ($\cal{S_{\rm tot}}$) and neglecting ($\cal{S_{\rm stat}}$) the systematic uncertainties.}
\begin{tabular}{c|c|c|c|c|c|c|c|c}
\multicolumn{9}{c}{$e^+e^- \rightarrow \eta_{\rm c} \pi^+\pi^-\pi^0$}\\ \hline
$\sqrt{s}$ [GeV] & ${\cal L}$ [pb$^{-1}$] & $N_{\rm obs}$ & $\kappa$ & $f_{\rm VP}$ & $\sum\varepsilon_i{\cal B}_i$ [\%] & $\sigma_{\rm Born}$ [pb] & UL$_{\rm 90}$ [pb] & ${\cal S}_{\rm stat}/{\cal S}_{\rm tot}$ [$\sigma$] \\ \hline
4.1780  &  3189.0  &  $530 \pm 246$  &  ~~~[0.720, 0.734]~~~  &  1.056  &  2.0  &  $10.4\;^{+  5.0}_{-  4.9} \pm  2.9$  & 17.9  &  2.2 / 1.9  \\
4.2263  &  1091.7  &  $786 \pm 159$  &  [0.716, 0.731]  &  1.056  &  2.0  &  $46.1\;^{+  9.5}_{-  9.4} \pm  6.6$  & 61.0  &  5.1 / 4.6  \\
4.2580  &   825.7  &  $465 \pm 134$  &  [0.786, 0.824]  &  1.054  &  2.0  &  $31.4\;^{+  9.6}_{-  9.6} \pm  6.7$  & 46.6  &  3.5 / 3.2  \\
4.3583  &   539.8  &  $242 \pm 115$  &  [0.802, 0.880]  &  1.051  &  2.1  &  $22.2\;^{+ 11.4}_{- 11.3} \pm  6.2$  & 39.2  &  2.2 / 1.9  \\
4.4156  &  1073.6  &  $379 \pm 165$  &  [0.780, 0.850]  &  1.053  &  2.2  &  $18.1\;^{+  8.4}_{-  8.4} \pm  4.5$  & 30.6  &  2.3 / 2.1  \\
4.5995  &   566.9  &  $ 79 \pm 102$  &  [0.763, 0.807]  &  1.055  &  2.0  &  $ 7.4\;^{+ 10.6}_{- 10.5} \pm  3.9$  & 23.9  &  0.8 / 0.7  \\
\end{tabular}  
\end{center}
\end{table*}
\renewcommand{\arraystretch}{1.0}

In conclusion, the process $e^+e^-\to \eta_{\rm c}\pi^+\pi^-\pi^0$ is observed for the first time. Furthermore, the measured energy-dependent cross section is found in agreement with the intermediate production of the $Y(4260)$ resonance, decaying to the $\eta_{\rm c}\pi^+\pi^-\pi^0$ final state.
%
%
%
\subsection{\label{subsec:etacULS} Upper limits on $\sigma_{\rm B}(e^+e^-\to\eta_{\rm c}\pi^+\pi^-)$ and $\sigma_{\rm B}(e^+e^-\to \eta_{\rm c}\pi^0\gamma)$}
\begin{figure}[tp!]
    \begin{center}
     \includegraphics[clip,trim= 0 0 0 0, width=0.49\linewidth, angle=0]{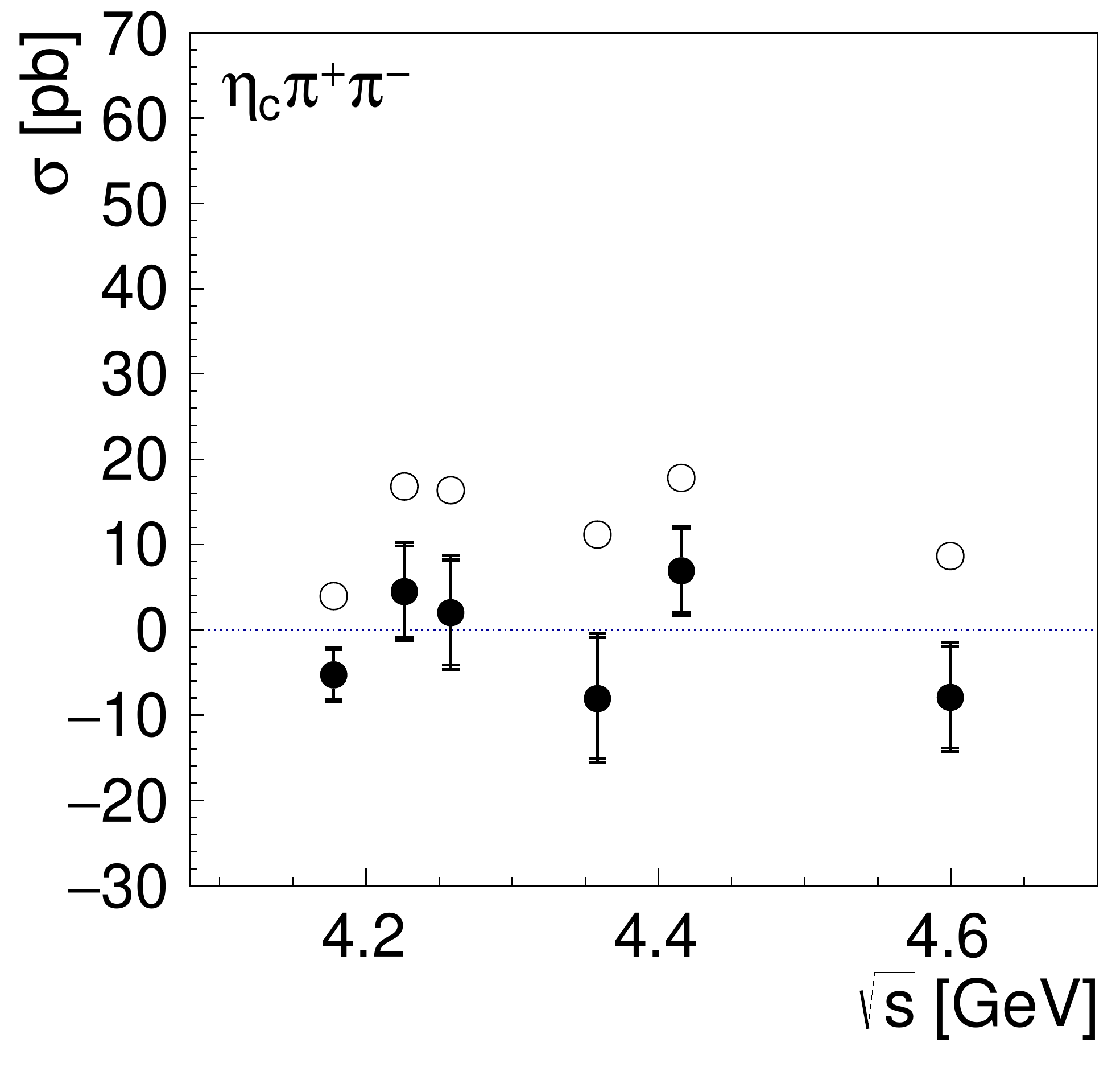}
     \includegraphics[clip,trim= 0 0 0 0, width=0.49\linewidth, angle=0]{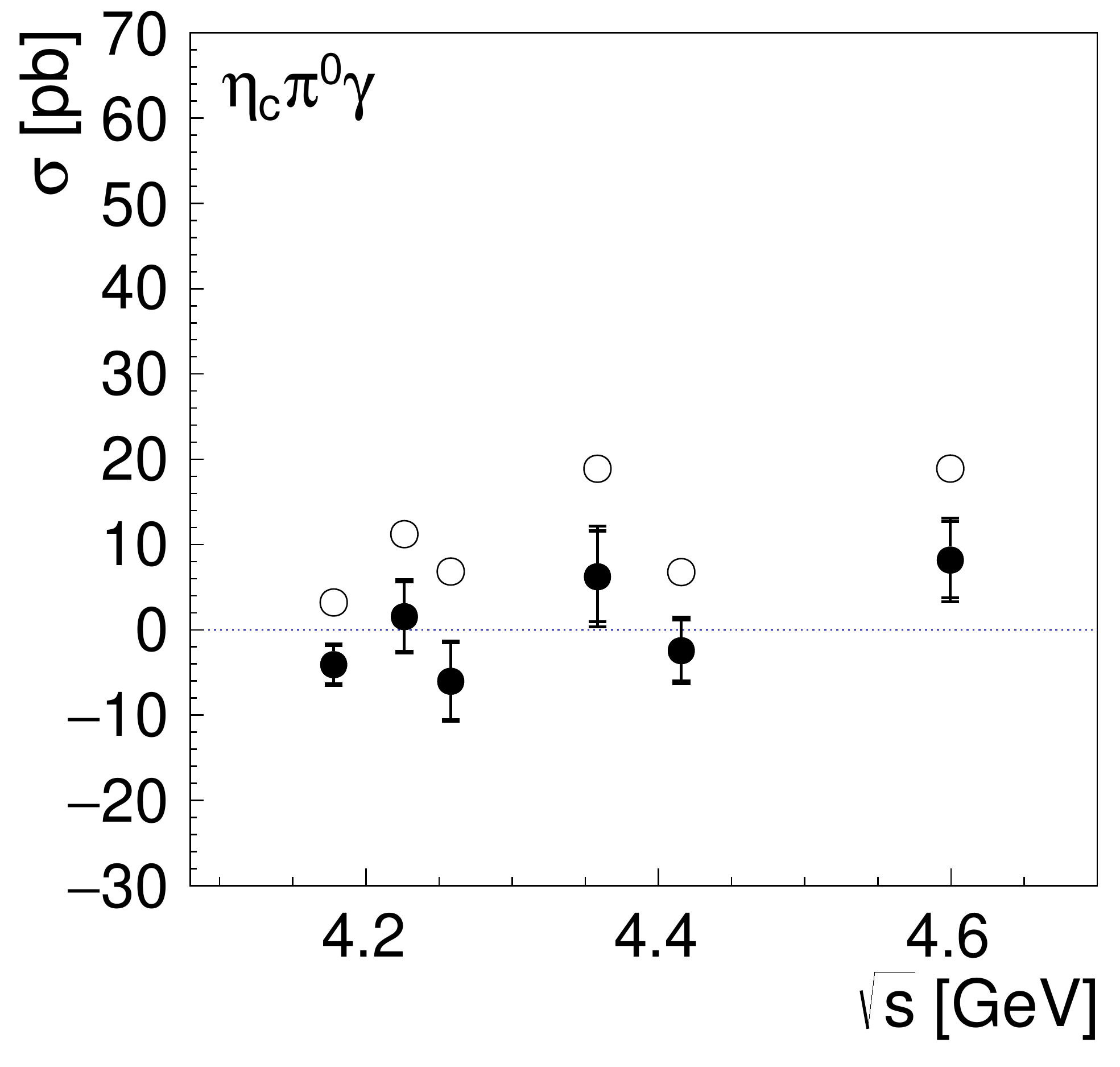}
     \caption{Measurements of the production cross section of $e^+e^-\to\eta_{\rm c}\pi^+\pi^-$ ({\it left}) and $e^+e^-\to\eta_{\rm c}\pi^0\gamma$ ({\it right}). No significant $\eta_{\rm c}$ production is observed for both recoil production modes over the covered $\sqrt{s}$ range from 4.18\,GeV to 4.60\,GeV. All the directly observed cross sections $\sigma$ are compatible with zero. The corresponding 90\,\% C.L. upper limits on the Born cross section (open circles) are given for all measured $\sigma(\sqrt{s})$.}
      \label{fig:EtacResults_etac_M7u9}
     \end{center}
\end{figure}

The cross section measurements of the other two production modes of $\eta_{\rm c}$ plus recoil particles, namely  $e^+e^-\to\eta_{\rm c}\pi^+\pi^-$ and $e^+e^-\to \eta_{\rm c}\pi^0\gamma$, are summarized in Tabs.\,\ref{tab:ResultsM7} and \ref{tab:ResultsM9}, respectively. As it can  be seen from Fig.\,\ref{fig:EtacResults_etac_M7u9}, where both results are graphically summarized, the measured cross sections $\sigma$ are compatible with zero.
%
%

The values of the directly measured cross sections $\sigma$, {\it i.e.} without radiative corrections applied, are shown (black data points in Fig.\,\ref{fig:EtacResults_etac_M7u9}). For the calculation of the upper limits on the Born cross sections (open circles in Fig.\,\ref{fig:EtacResults_etac_M7u9}), the conservative procedure of radiative corrections has been applied. This is based on the global minimum $\kappa_{\rm min}$ value determined under the assumption of a narrow resonance, as explained in Sec.\,\ref{subsec:RadCorr}. 

The resulting UL$_{\rm 90}$ values range from about 3 to 4\,pb up to about 19 to 18\,pb for the different $\sqrt{s}$ data sets, respectively. All relevant numbers and results for these two measurements as well as the applied $\kappa_{\rm min}$ values are summarized in Tab.\,\ref{tab:ResultsM7} and Tab.\,\ref{tab:ResultsM9}.
\renewcommand{\arraystretch}{1.5}
\begin{table*}[h]
\begin{center}
  \caption{\label{tab:ResultsM7} 
    Summary of production cross section results for $e^+e^- \rightarrow \eta_{\rm c} \pi^+\pi^-$ based on the six 
    different center-of-mass energy $\sqrt{s}$ data sets~\cite{bes3_ecms_paper} of integrated luminosities 
    $\cal{L}$~\cite{bes3_lumi_paper}. Quoted are the number of observed events $N_{\rm obs}$ as obtained from 
    the simultaneous fits, the applied radiative correction factor $\kappa_{\rm min}$, the applied 
    vacuum-polarization factor $f_{\rm VP}$, the sum of the 16 efficiencies times branching ratio values 
    $\sum\varepsilon_i{\cal B}_i$, the Born cross section $\sigma_{\rm Born}$ as well as the computed 
    90\,\% C.L. upper limits UL$_{\rm 90}$, and the computed statistical significance including 
    ($\cal{S_{\rm tot}}$) and neglecting ($\cal{S_{\rm stat}}$) the systematic uncertainties.}
\begin{tabular}{c|c|c|c|c|c|c|c|c}
\multicolumn{9}{c}{$e^+e^- \rightarrow \eta_{\rm c} \pi^+\pi^-$}\\ \hline
$\sqrt{s}$ [GeV] & ${\cal L}$ [pb$^{-1}$] & $N_{\rm obs}$ & $\kappa_{\rm min}$ & $f_{\rm VP}$ & $\sum\varepsilon_i{\cal B}_i$ [\%] & $\sigma_{\rm Born}$ [pb] & UL$_{\rm 90}$ [pb] & ${\cal S}_{\rm stat}/{\cal S}_{\rm tot}$ [$\sigma$] \\ \hline
4.1780 &  3189.0 & $-768 \pm 413$  &  ~~~0.628~~~ &  1.056  &  4.4  &   $ -7.2\;^{+  4.0}_{-  4.0} \pm  1.5$  &   4.0  &  0.0 / 0.0  \\
4.2263 &  1091.7 & $ 197 \pm 241$  &  0.627 &  1.056  &  4.2  &   $  6.0\;^{+  7.1}_{-  7.1} \pm  2.6$  &  16.8  &  0.8 / 0.8  \\
4.2580 &  825.7  & $  75 \pm 209$  &  0.627 &  1.054  &  4.1  &   $  2.8\;^{+  8.1}_{-  8.1} \pm  3.4$  &  16.4  &  0.4 / 0.3  \\
4.3583 &  539.8  & $-162 \pm 152$  &  0.626 &  1.051  &  4.0  &   $-10.5\;^{+  9.3}_{-  9.2} \pm  3.1$  &  11.2  &  0.0 / 0.0  \\
4.4156 &  1073.6 & $ 278 \pm 201$  &  0.625 &  1.053  &  3.8  &   $  8.9\;^{+  6.2}_{-  6.1} \pm  2.3$  &  17.8  &  1.5 / 1.4  \\
4.5995 &  566.9  &  $-152 \pm 121$ &  0.624 &  1.055  &  3.6  &   $-10.1\;^{+  7.5}_{-  7.4} \pm  2.9$  &   8.7  &  0.0 / 0.0  \\
\end{tabular}  
\end{center}
\end{table*}
\renewcommand{\arraystretch}{1.0}
\renewcommand{\arraystretch}{1.5}
\begin{table*}[h]
\begin{center}
  \caption{\label{tab:ResultsM9} 
    Summary of production cross section results for $e^+e^- \rightarrow \eta_{\rm c} \pi^0\gamma$ based on the six different 
    center-of-mass energy $\sqrt{s}$ data sets~\cite{bes3_ecms_paper} of integrated luminosities $\cal{L}$~\cite{bes3_lumi_paper}. 
    Quoted are the number of observed events $N_{\rm obs}$ as obtained from the simultaneous fits, the applied $\kappa_{\rm min}$, 
    the applied vacuum-polarization factor $f_{\rm VP}$, the sum of the 16 efficiencies times branching ratio values 
    $\sum\varepsilon_i{\cal B}_i$, the Born cross section $\sigma_{\rm Born}$ as well as the computed 90\,\% C.L. upper limits 
    UL$_{\rm 90}$, and the computed significance including ($\cal{S_{\rm tot}}$) and neglecting ($\cal{S_{\rm stat}}$) the systematic 
    uncertainties.}
\begin{tabular}{c|c|c|c|c|c|c|c|c}
\multicolumn{9}{c}{$e^+e^- \rightarrow \eta_{\rm c} \pi^0\gamma$}\\ \hline
$\sqrt{s}$ [GeV] & ${\cal L}$ [pb$^{-1}$] & $N_{\rm obs}$ & $\kappa_{\rm min}$ & $f_{\rm VP}$ & $\sum\varepsilon_i{\cal B}_i$ [\%] & $\sigma_{\rm Born}$ [pb] & UL$_{\rm 90}$ [pb] & ${\cal S}_{\rm stat}/{\cal S}_{\rm tot}$ [$\sigma$] \\ \hline
4.1780 &  3189.0 &  $-378 \pm 216$ & ~~~0.628~~~ & 1.056 &  3.0 & $ -5.7\;^{+  3.2}_{-  3.2} \pm  1.2 $ &   3.2  &  0.0 / 0.0  \\
4.2263 &  1091.7 &  $  63 \pm 125$ & 0.627 & 1.056 &  2.8 & $  2.1\;^{+  5.6}_{-  5.6} \pm  1.8 $  &  11.2  &  0.5 / 0.4  \\
4.2580 &   825.7 &  $-125 \pm 106$ & 0.627 & 1.054 &  2.8 & $ -8.2\;^{+  6.1}_{-  6.1} \pm  2.2 $ &   7.0  &  0.0 / 0.0  \\
4.3583 &   539.8 &  $  92 \pm  81$ & 0.626 & 1.051 &  2.8 & $  7.9\;^{+  7.1}_{-  7.0} \pm  2.2 $  &  18.3  &  1.2 / 1.1  \\
4.4156 &  1073.6 &  $ -58 \pm 107$ & 0.625 & 1.053 &  2.8 & $ -3.1\;^{+  4.7}_{-  4.7} \pm  1.5 $ &   6.6  &  0.0 / 0.0  \\
4.5995 &  566.9  &  $ 140 \pm  72$ & 0.625 & 1.055 &  2.8 & $ 10.6\;^{+  5.8}_{-  5.7} \pm  2.3 $  &  18.9  &  2.1 / 1.8  \\
\end{tabular}  
\end{center}
\end{table*}
\renewcommand{\arraystretch}{1.0}
%
%
\begin{figure*}[tp!]
    \begin{center}
     \includegraphics[clip,trim= 0 0 0 0, width=1.0\linewidth, angle=0]{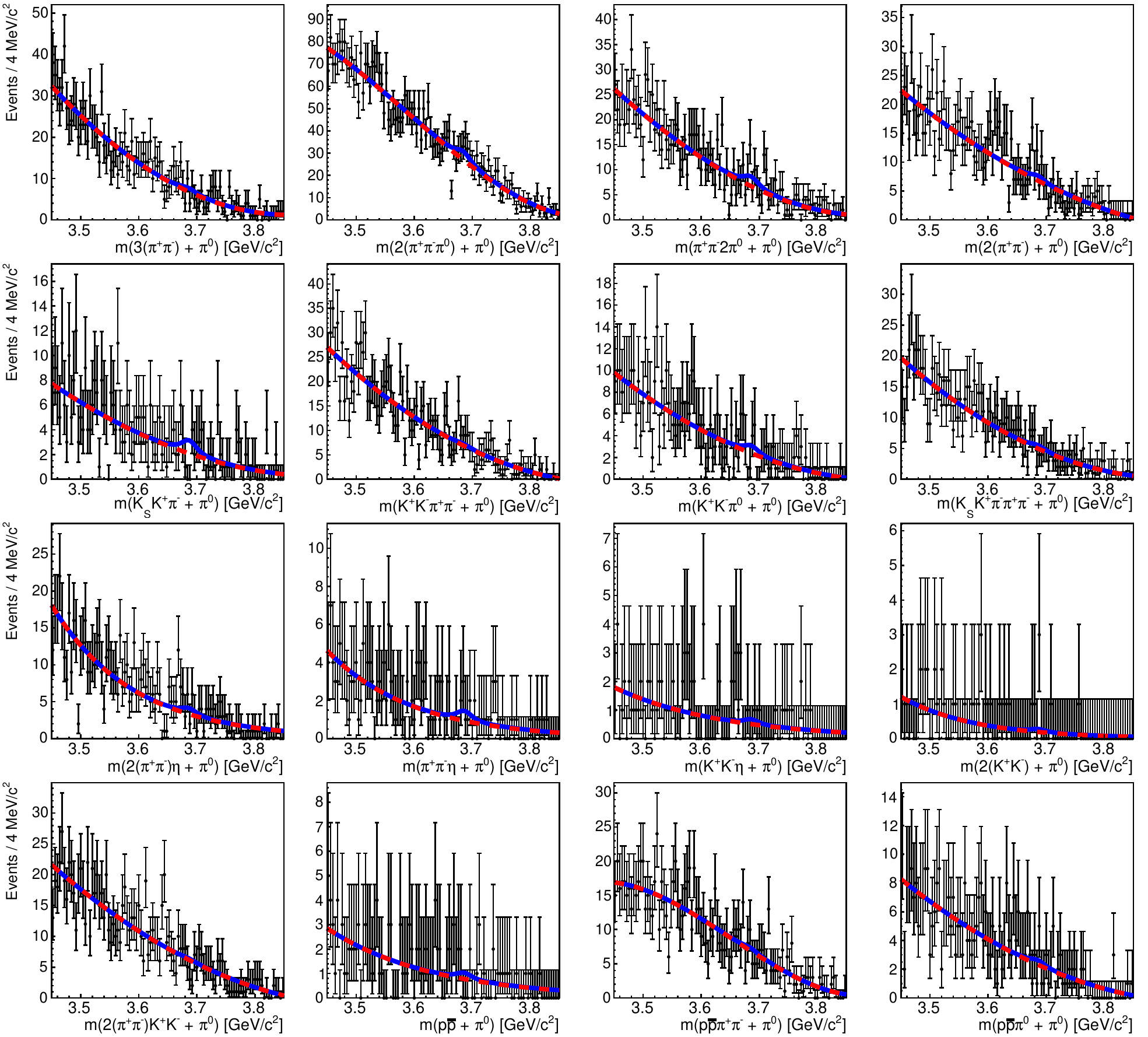}
     \caption{Simultaneous fit result for the $Z_{\rm c}$ search, with radiative corrections included, for the example case of 
       $e^+e^- \to Z_{\rm c}^{0}  \pi^{+}\pi^{-} \to (\eta_{\rm c} \pi^0) \pi^+ \pi^-$ at $\sqrt{s}=4.23\,$GeV, shown here for the 
       assumed $Z_{\rm c}$ parameters of $(m_{\rm Z_{\rm c}},\Gamma_{\rm Z_{\rm c}}) = (3685\,{\rm MeV}/c^2, 28\,{\rm MeV})$. 
       The mass spectra are shown for the 16 underlying $\eta_{\rm c}$ decay channels together with the simultaneous fit (blue solid curve).}
      \label{fig:simuFit_Zc_M4_m3685_w28}
     \end{center}
\end{figure*}
%
%
\begin{figure*}[tp!]
    \begin{center}
     \includegraphics[clip,trim= 0 0 0 5, width=0.95\linewidth, angle=0]{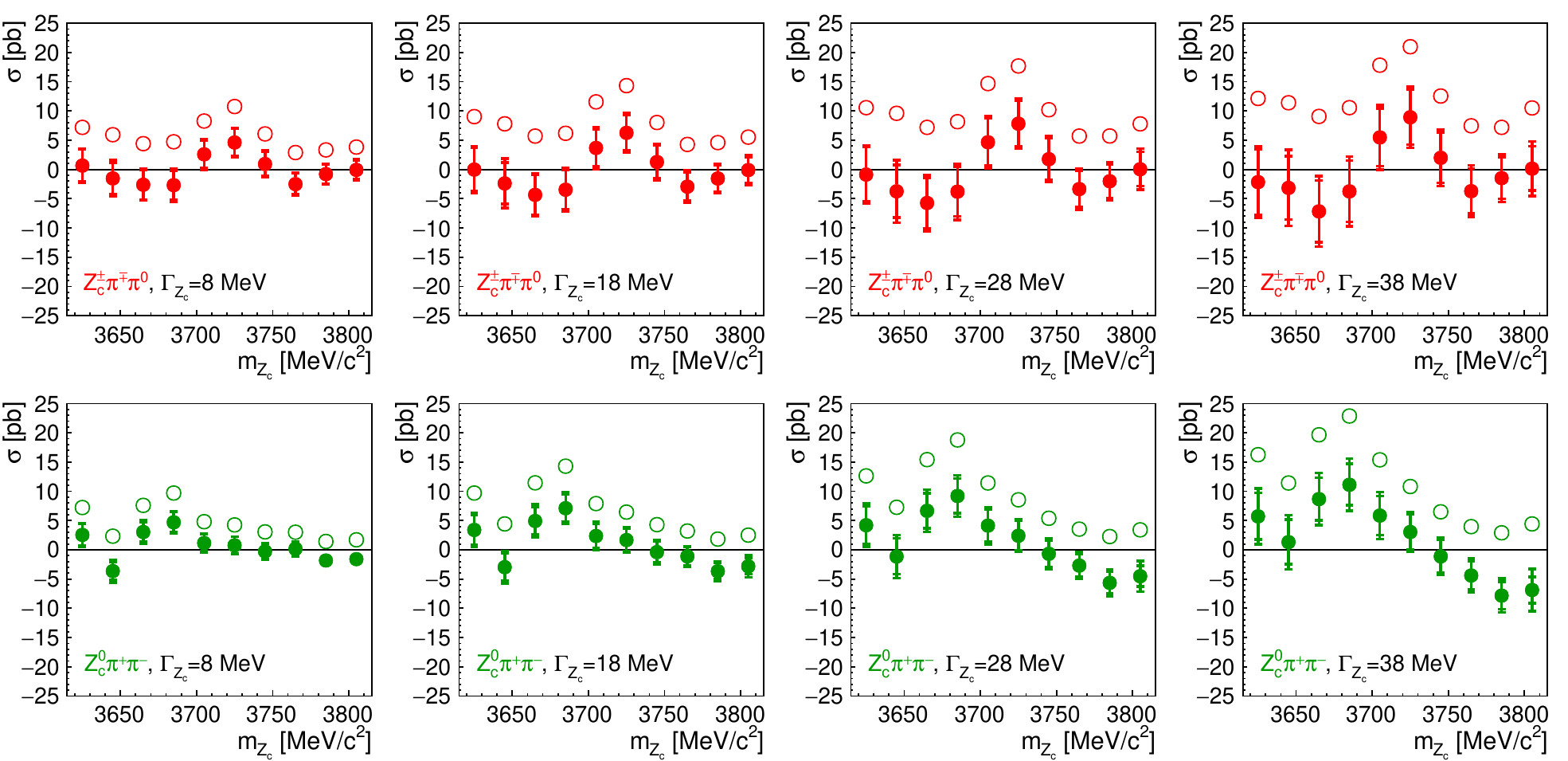}
     \caption{
       Search for a possible charged ({\it top}) or neutral ({\it bottom}) $Z_{\rm c}$ state in the vicinity of the $D\bar{D}$ threshold, 
       decaying to $\eta_{\rm c}\pi$ in $e^+e^- \to \eta_{\rm c} \pi^0 \pi^+ \pi^-$ reactions.  
       Shown are the results of mass and width scans in terms of the measured $(m_{\rm Z_{\rm c}},\Gamma_{\rm Z_{\rm c}})$-dependent 
       cross section $\sigma$ (red/green data points) together with the upper limits (open circles) on the Born cross section $\sigma_{\rm Born}$.
       The measurements have been performed for  four assumed widths $\Gamma_{\rm Zc}=8, 18, 28, 38$\,MeV 
       and ten assumed masses $m_{\rm Z_{\rm c}} = 3625, 3645, 3665, 3685, 3705, 3725, 3745, 3765, 3785, 3805$\,MeV/$c^2$ at $\sqrt{s}= 4.23$\,GeV.
}
      \label{fig:ZcSearchResults}
     \end{center}
\end{figure*}   

%
%
\section{\label{sec:Zc} Search for $Z_{\rm c} \to \eta_{\rm c}\pi$}

We perform a search for a possible $Z_{\rm c}$ state in the vicinity of the $D\bar{D}$ threshold, decaying to $\eta_{\rm c}\pi$. Given the non-observation of significant $e^+e^-$ productions of $\eta_{\rm c}\pi^+\pi^-$ or $\eta_{\rm c}\pi^0\gamma$ (Sec.\,\ref{subsec:etacULS}) and the observation of an (underlying) $\eta_{\rm c}$ production cross section for $e^+e^-\to \eta_{\rm c}\pi^+\pi^-\pi^0$ with a statistical significance of about $5\sigma$ (Sec.\,\ref{subsec:etacpipipi0}), we restrict our $Z_{\rm c}$ search to the latter final state.

Concretely, we have searched for a charged or a neutral $Z_{\rm c}^{\pm/0}$ decaying to $\eta_{\rm c} \pi^{\pm/0}$ in the two reaction modes
\begin{itemize}
\item[a)] $e^+e^- \to Z_{\rm c}^{+} \pi^{-}\pi^{0} \to (\eta_{\rm c} \pi^+) \pi^-\pi^0$ (or c.c.), and 
\item[b)] $e^+e^- \to Z_{\rm c}^{0}  \pi^{+}\pi^{-} \to (\eta_{\rm c} \pi^0) \pi^+ \pi^-$. 
\end{itemize}
The search in both modes is based just on the data set taken at $\sqrt{s} = 4.23$ GeV, where the underlying $\eta_{\rm c}$ production has been measured to be most significant (Tab.\,\ref{tab:ResultsM8}). Since such a state has not been observed yet, the search is realized in terms of a mass and a width scan, testing in total $10\times 4 =40$ different $(m,\Gamma)$ parameter combinations per charge mode. As an example, a simultaneous fit to the corresponding invariant mass spectra is shown in Fig.\,\ref{fig:simuFit_Zc_M4_m3685_w28}.

Inspired by already observed $Z_{\rm c}$ states, corresponding signal MC data samples are generated and reconstructed ({\it cf.} Secs.\,\ref{sec:DetsData} and \ref{subsec:EvtSel}) for an assumed natural decay width of $\Gamma_{\rm Z_{\rm c}}=28\,$MeV and four different assumed $Z_{\rm c}$ masses $m_{\rm Z_{\rm c}}=3645, 3685, 3745, 3805\;$MeV/$c^2$. The reconstruction efficiencies show a shallow linear dependence on the $Z_{\rm c}$ mass assumed in the full signal MC simulation and are (linearly) interpolated for intermediate parameter settings. The $Z_{\rm c}$ signal line shapes for the 16 different $\eta_{\rm c}$ decay modes are used as obtained from the simulated signal MC simulations for $m_{\rm Z_{\rm c}}=3745\,$MeV/$c^2$.

The $Z_{\rm c}$ mass $m_{\rm Z_{\rm c}}$ is varied in 20\,MeV/$c^2$ steps in the $m_{\rm Z_{\rm c}}$ range of $[3625,3805]$\,MeV/$c^2$ (by fixing the mass parameter in the Voigtian function that represents the signal, {\it cf.} Sec.\,\ref{subsec:RecoEffi}). Similarly, the scan of the width $\Gamma_{\rm Z_{\rm c}}$ is varied in 10\,MeV steps, $\Gamma_{\rm Z_{\rm c}}= 8, 18, 28, 38\,$MeV. Finally, we perform for each mass and width combination $(m_{\rm Z_{\rm c}},\Gamma_{\rm Z_{\rm c}})$ the full analysis in terms of a simultaneous fit, as was done for the $\eta_{\rm c}$ cases (Sec.\,\ref{sec:Xsec}), and extract the $(m_{\rm Z_{\rm c}},\Gamma_{\rm Z_{\rm c}})$-dependent cross section $\sigma$.

The outcome of the mass-width scan is graphically summarized in Fig.\,\ref{fig:ZcSearchResults}. The measured cross sections, calculated according to Eq.\,(\ref{Eq.simuFit}), for the four different assumed widths $\Gamma_{\rm Z_{\rm c}}$ are plotted versus the ten assumed masses $m_{\rm Z_{\rm c}}$, for both a possible charged and a possible neutral $Z_{\rm c}$ state, resulting in a total number of 80 individual measurements. 

No clear signal is found for a charged $Z_{\rm c}^\pm \to \eta_{\rm c}\pi^\pm$ state. Here, the measured cross sections are statistically consistent with zero. The corresponding statistical significances of a possible signal for the different $(m_{\rm Z_{\rm c}},\Gamma_{\rm Z_{\rm c}})$ combinations are  mostly well below $1\sigma$, except from the assumed masses $m_{\rm Z_{\rm c}}$ of 3705\,MeV/$c^2$ and 3725\,MeV/$c^2$, for which significances of up to $2\sigma$ are found. 

The measured cross sections for a possible neutral $Z_{\rm c}^0 \to \eta_{\rm c}\pi^0$ state are found to be more significant. In particular, for the assumed $m_{\rm Z_{\rm c}}= 3685$\,MeV/$c^2$, statistical significances of about $3\sigma$ are found for all four assumed widths $\Gamma_{\rm Z_{\rm c}}$. The maximum statistical significance observed is ${\cal S}_{\rm stat}= 3.2\sigma$, whereas taking into account the systematic uncertainties (those marked with a star in Tab.\,\ref{tab:RankingSystError_Etac_M8_4230}), the resultant significance is still ${\cal S}_{\rm tot}=2.8\sigma$.

Applying the conservative approach of radiative corrections (Sec.\,\ref{subsec:RadCorr}), also UL$_{90}$ values on the Born cross sections are provided (open circles in Fig.\,\ref{fig:ZcSearchResults}). All relevant absolute numbers of these results are also summarized in Tabs.\,\ref{tab:ResultsM3} and \ref{tab:ResultsM4}, respectively, and the corresponding statistical significances are listed in Tabs.\,\ref{tab:ResultsSignificancesM3} and \ref{tab:ResultsSignificancesM4}.  

The applied mass-width scan method has been validated by a blind mixed-in signal check, and the mass-width resolution obtained by the chosen scan points has been validated to be sufficient as well. 

It is necessary to correct the width and mass scan for the so-called ``Look-elsewhere'' effect~\cite{LookElseWhereEffect}. We have studied this effect by random generation of (about $ 8\cdot 10^5$ times) the 16 mass spectra for the different final states according to the background distributions obtained by fits to the data. After simultaneous fits of the $10 \times 4 = 40$ fixed $(m, \Gamma)$ combinations, we performed a likelihood-ratio test to determine the maximum of the 40 fitted likelihood ratios $LR_{\rm max} = \max_{(\rm m,\Gamma)} 2 \cdot \ln (L_1 / L_0)$, where $L_1$ is the likelihood-value of the fit including a possible signal and $L_0$ the one for the null-hypothesis associated with no signal. The fraction of cases with $LR_{\rm max}>LR_{\rm data}$ (with $LR_{\rm data} = S_{\rm data}^2$ and $S_{\rm data} = 3.2\sigma$ being our largest observed significance) represents an estimate for the true $p$-value that corresponds to a reduced significance of about $S=2\sigma$ for the resonance parameters of $m_{\rm Z_{\rm c}} = 3685\,{\rm MeV}/c^2$ and $\Gamma_{\rm Z_{\rm c}}= 28\,{\rm MeV}$.
\renewcommand{\arraystretch}{1.5}
\begin{table*}[h]
\begin{center}
  \caption{\label{tab:ResultsM3}
    Summary of results for the $Z_c^\pm \rightarrow \eta_{\rm c}\pi^\pm$ at $\sqrt{s}= 4.23$\,GeV. 
    Quoted are the measured production cross sections $\sigma$ as obtained from the simultaneous fits, 
    without radiative corrections applied. 
    The measured values assigned with the statistical and systematic uncertainties are followed by the 
    90\,\% C.L. upper limit values UL$_{90}$ on the Born cross section $\sigma_{\rm Born}$ in brackets, both in units of [pb].}
\begin{tabular}{c|r@{\hspace{2mm}}r|r@{\hspace{2mm}}r|r@{\hspace{2mm}}r|r@{\hspace{2mm}}r}
  $m_{Z_{\rm c}}$ [GeV/$c^{2}$]  &  \multicolumn{2}{|c|}{$\Gamma = 8$ MeV}  &  \multicolumn{2}{|c|}{$\Gamma = 18$ MeV}  &  \multicolumn{2}{|c|}{$\Gamma = 28$ MeV}  &  \multicolumn{2}{|c}{$\Gamma = 38$ MeV} \\ \hline
  3625  &  $  0.7\;^{+  2.8}_{-  2.8} \pm  0.9$ &  (7.2) &  $ -0.0\;^{+  3.8}_{-  3.7} \pm  1.4$ &  (9.0) &  $ -0.9\;^{+  4.7}_{-  4.6} \pm  1.9$ & (10.6) &  $ -2.2\;^{+  5.6}_{-  5.6} \pm  2.4$ & (12.1)\\ 
  3645  &  $ -1.5\;^{+  2.7}_{-  2.6} \pm  1.7$ &  (5.9) &  $ -2.4\;^{+  3.6}_{-  3.5} \pm  2.3$ &  (7.8) &  $ -3.7\;^{+  4.5}_{-  4.5} \pm  2.9$ &  (9.6) &  $ -3.2\;^{+  5.5}_{-  5.4} \pm  3.6$ & (11.4)\\ 
  3665  &  $ -2.6\;^{+  2.6}_{-  2.5} \pm  1.0$ &  (4.4) &  $ -4.3\;^{+  3.4}_{-  3.4} \pm  1.5$ &  (5.7) &  $ -5.7\;^{+  4.4}_{-  4.3} \pm  2.2$ &  (7.2) &  $ -7.2\;^{+  5.3}_{-  5.3} \pm  2.9$ &  (9.1)\\ 
  3685  &  $ -2.7\;^{+  2.5}_{-  2.5} \pm  1.5$ &  (4.7) &  $ -3.5\;^{+  3.4}_{-  3.3} \pm  1.7$ &  (6.2) &  $ -3.8\;^{+  4.3}_{-  4.2} \pm  2.3$ &  (8.2) &  $ -3.8\;^{+  5.3}_{-  5.2} \pm  2.9$ & (10.6)\\ 
  3705  &  $  2.6\;^{+  2.4}_{-  2.4} \pm  1.1$ &  (8.3) &  $  3.7\;^{+  3.2}_{-  3.2} \pm  1.6$ & (11.6) &  $  4.7\;^{+  4.1}_{-  4.0} \pm  2.0$ & (14.7) &  $  5.5\;^{+  5.0}_{-  4.9} \pm  2.5$ & (17.8)\\ 
  3725  &  $  4.6\;^{+  2.3}_{-  2.3} \pm  1.0$ & (10.8) &  $  6.3\;^{+  3.1}_{-  3.1} \pm  1.4$ & (14.3) &  $  7.8\;^{+  3.9}_{-  3.8} \pm  1.8$ & (17.7) &  $  8.9\;^{+  4.7}_{-  4.7} \pm  2.3$ & (21.0)\\ 
  3745  &  $  0.9\;^{+  2.1}_{-  2.0} \pm  1.0$ &  (6.1) &  $  1.3\;^{+  2.8}_{-  2.8} \pm  1.3$ &  (8.0) &  $  1.8\;^{+  3.6}_{-  3.5} \pm  1.7$ & (10.2) &  $  2.0\;^{+  4.4}_{-  4.3} \pm  2.2$ & (12.6)\\ 
  3765  &  $ -2.5\;^{+  1.8}_{-  1.8} \pm  0.9$ &  (2.9) &  $ -2.9\;^{+  2.5}_{-  2.4} \pm  1.2$ &  (4.3) &  $ -3.4\;^{+  3.2}_{-  3.1} \pm  1.6$ &  (5.7) &  $ -3.7\;^{+  3.9}_{-  3.8} \pm  2.1$ &  (7.5)\\ 
  3785  &  $ -0.8\;^{+  1.7}_{-  1.6} \pm  0.7$ &  (3.3) &  $ -1.5\;^{+  2.3}_{-  2.2} \pm  1.1$ &  (4.6) &  $ -2.0\;^{+  2.9}_{-  2.9} \pm  1.6$ &  (5.7) &  $ -1.5\;^{+  3.6}_{-  3.6} \pm  2.0$ &  (7.2)\\ 
  3805  &  $ -0.1\;^{+  1.6}_{-  1.5} \pm  0.9$ &  (3.8) &  $ -0.1\;^{+  2.2}_{-  2.2} \pm  1.4$ &  (5.5) &  $  0.0\;^{+  3.0}_{-  2.9} \pm  2.0$ &  (7.8) &  $  0.1\;^{+  3.8}_{-  3.7} \pm  2.8$ & (10.5)\\ 
\end{tabular}
\end{center}
\end{table*}
\renewcommand{\arraystretch}{1.0}
\renewcommand{\arraystretch}{1.5}
\begin{table*}[h]
\begin{center}
  \caption{\label{tab:ResultsM4}
    Summary of results for the $Z_c^0 \rightarrow \eta_{\rm c}\pi^0$ at $\sqrt{s}= 4.23$\,GeV. 
    Quoted are the measured production cross sections $\sigma$ as obtained from the simultaneous fits, 
    without radiative corrections applied.
    The measured values assigned with the statistical and systematic uncertainties are followed by the 
    90\,\% C.L. upper limit values UL$_{90}$ on the Born cross section $\sigma_{\rm Born}$ in brackets, both in units of [pb].
  }
\begin{tabular}{c|r@{\hspace{2mm}}r|r@{\hspace{2mm}}r|r@{\hspace{2mm}}r|r@{\hspace{2mm}}r}
  $m_{Z_{\rm c}}$ [GeV/$c^{2}$]  &  \multicolumn{2}{|c|}{$\Gamma = 8$ MeV}  &  \multicolumn{2}{|c|}{$\Gamma = 18$ MeV}  &  \multicolumn{2}{|c|}{$\Gamma = 28$ MeV}  &  \multicolumn{2}{|c}{$\Gamma = 38$ MeV} \\ \hline
  3625  &  $  2.6\;^{+  1.9}_{-  1.9} \pm  0.8$ & (7.2) &  $  3.4\;^{+  2.6}_{-  2.5} \pm  1.3$ &  (9.7) &  $  4.2\;^{+  3.3}_{-  3.2} \pm  1.9$ & (12.6) &  $  5.7\;^{+  4.0}_{-  3.9} \pm  2.7$ & (16.3)\\ 
  3645  &  $ -3.6\;^{+  1.6}_{-  1.5} \pm  1.2$ & (2.3) &  $ -3.0\;^{+  2.4}_{-  2.3} \pm  1.5$ &  (4.4) &  $ -1.1\;^{+  3.1}_{-  3.0} \pm  2.1$ &  (7.3) &  $  1.3\;^{+  3.8}_{-  3.8} \pm  2.7$ & (11.4)\\ 
  3665  &  $  3.0\;^{+  1.7}_{-  1.7} \pm  1.1$ & (7.6) &  $  4.9\;^{+  2.4}_{-  2.3} \pm  1.6$ & (11.5) &  $  6.7\;^{+  3.0}_{-  3.0} \pm  2.0$ & (15.4) &  $  8.7\;^{+  3.7}_{-  3.6} \pm  2.6$ & (19.7)\\ 
  3685  &  $  4.7\;^{+  1.7}_{-  1.7} \pm  0.9$ & (9.7) &  $  7.1\;^{+  2.4}_{-  2.3} \pm  1.4$ & (14.3) &  $  9.2\;^{+  3.0}_{-  2.9} \pm  2.0$ & (18.8) &  $ 11.1\;^{+  3.6}_{-  3.5} \pm  2.7$ & (22.9)\\ 
  3705  &  $  1.2\;^{+  1.5}_{-  1.4} \pm  0.8$ & (4.8) &  $  2.4\;^{+  2.1}_{-  2.1} \pm  1.2$ &  (7.9) &  $  4.1\;^{+  2.8}_{-  2.7} \pm  1.7$ & (11.4) &  $  5.9\;^{+  3.4}_{-  3.3} \pm  2.2$ & (15.4)\\ 
  3725  &  $  0.8\;^{+  1.5}_{-  1.4} \pm  0.5$ & (4.3) &  $  1.7\;^{+  2.0}_{-  2.0} \pm  0.8$ &  (6.5) &  $  2.4\;^{+  2.6}_{-  2.5} \pm  1.2$ &  (8.6) &  $  3.0\;^{+  3.1}_{-  3.1} \pm  1.5$ & (10.8)\\ 
  3745  &  $ -0.3\;^{+  1.3}_{-  1.3} \pm  0.7$ & (3.1) &  $ -0.4\;^{+  1.8}_{-  1.8} \pm  1.0$ &  (4.3) &  $ -0.7\;^{+  2.3}_{-  2.3} \pm  1.3$ &  (5.4) &  $ -1.1\;^{+  2.9}_{-  2.8} \pm  1.5$ &  (6.5)\\ 
  3765  &  $  0.2\;^{+  1.1}_{-  1.1} \pm  0.6$ & (3.0) &  $ -1.1\;^{+  1.6}_{-  1.5} \pm  0.9$ &  (3.2) &  $ -2.7\;^{+  2.0}_{-  2.0} \pm  1.2$ &  (3.5) &  $ -4.4\;^{+  2.5}_{-  2.4} \pm  1.5$ &  (4.0)\\ 
  3785  &  $ -1.8\;^{+  1.0}_{-  0.9} \pm  0.6$ & (1.4) &  $ -3.7\;^{+  1.4}_{-  1.4} \pm  0.9$ &  (1.8) &  $ -5.6\;^{+  1.9}_{-  1.8} \pm  1.3$ &  (2.3) &  $ -7.8\;^{+  2.4}_{-  2.3} \pm  1.8$ &  (2.9)\\ 
  3805  &  $ -1.6\;^{+  1.0}_{-  0.9} \pm  0.7$ & (1.7) &  $ -2.8\;^{+  1.4}_{-  1.3} \pm  1.3$ &  (2.5) &  $ -4.5\;^{+  1.8}_{-  1.8} \pm  2.0$ &  (3.4) &  $ -6.9\;^{+  2.3}_{-  2.2} \pm  2.8$ &  (4.4)\\ 
\end{tabular}
\end{center}
\end{table*}
\renewcommand{\arraystretch}{1.0}
\renewcommand{\arraystretch}{1.5}
\begin{table*}[h]
\begin{center}
  \caption{\label{tab:ResultsSignificancesM3} 
    Summary of signal significances $\cal{S_{\rm stat}}$ (purely statistical) and $\cal{S_{\rm tot}}$ (taking into account the systematic 
    uncertainties) for the performed $Z_{\rm c}^{\pm} \rightarrow \eta_{\rm c}\pi^\pm$ search at $\sqrt{s}= 4.23$\,GeV.
  }
\begin{tabular}{c|c|c|c|c}
  &  $\Gamma = 8$ MeV  &  $\Gamma = 18$ MeV  &  $\Gamma = 28$ MeV  &  $\Gamma = 38$ MeV \\ \hline
  $m_{\rm Z_c}$ [GeV/$c^{2}$]  &  ${\cal S}_{\rm stat} / {\cal S}_{\rm tot}$ [$\sigma$]  &  ${\cal S}_{\rm stat} / {\cal S}_{\rm tot}$ [$\sigma$]  &  ${\cal S}_{\rm stat} / {\cal S}_{\rm tot}$ [$\sigma$]  &  ${\cal S}_{\rm stat} / {\cal S}_{\rm tot}$ [$\sigma$] \\ \hline
  3625  &   0.1 /  0.2   &   0.0 /  0.0   &   0.0 /  0.0   &   0.0 /  0.0  \\ 
  3645  &   0.0 /  0.0   &   0.0 /  0.0   &   0.0 /  0.0   &   0.0 /  0.0  \\ 
  3665  &   0.0 /  0.0   &   0.0 /  0.0   &   0.0 /  0.0   &   0.0 /  0.0  \\ 
  3685  &   0.0 /  0.0   &   0.0 /  0.0   &   0.0 /  0.0   &   0.0 /  0.0  \\ 
  3705  &   1.0 /  1.0   &   1.0 /  1.1   &   1.0 /  1.1   &   1.0 /  1.0  \\ 
  3725  &   2.0 /  2.0   &   2.0 /  2.0   &   1.9 /  1.9   &   1.8 /  1.8  \\ 
  3745  &   0.4 /  0.4   &   0.4 /  0.5   &   0.5 /  0.5   &   0.5 /  0.4  \\ 
  3765  &   0.0 /  0.0   &   0.0 /  0.0   &   0.0 /  0.0   &   0.0 /  0.0  \\ 
  3785  &   0.0 /  0.0   &   0.0 /  0.0   &   0.0 /  0.0   &   0.0 /  0.0  \\ 
  3805  &   0.2 /  0.0   &   0.4 /  0.0   &   0.6 /  0.0   &   0.7 /  0.0  \\ 
\end{tabular}
\end{center}
\end{table*}
\renewcommand{\arraystretch}{1.0}
\renewcommand{\arraystretch}{1.5}
\begin{table*}[h]
\begin{center}
  \caption{\label{tab:ResultsSignificancesM4}
    Summary of signal significances $\cal{S_{\rm stat}}$ (purely statistical) and $\cal{S_{\rm tot}}$ (taking into account the systematic 
    uncertainties) for the performed $Z_{\rm c}^{0} \rightarrow \eta_{\rm c}\pi^0$ search at $\sqrt{s}= 4.23$\,GeV.
  }
\begin{tabular}{c|c|c|c|c}
  &  $\Gamma = 8$ MeV  &  $\Gamma = 18$ MeV  &  $\Gamma = 28$ MeV  &  $\Gamma = 38$ MeV \\ \hline
  $m_{\rm Z_c}$ [GeV/$c^{2}$] &  ${\cal S}_{\rm stat} / {\cal S}_{\rm tot}$ [$\sigma$]  &  ${\cal S}_{\rm stat} / {\cal S}_{\rm tot}$ [$\sigma$]  &  ${\cal S}_{\rm stat} / {\cal S}_{\rm tot}$ [$\sigma$]  &  ${\cal S}_{\rm stat} / {\cal S}_{\rm tot}$ [$\sigma$] \\ \hline
  3625  &   1.4 /  1.3   &   1.3 /  1.2   &   1.3 /  1.2   &   1.4 /  1.2  \\ 
  3645  &   0.0 /  0.0   &   0.0 /  0.0   &   0.0 /  0.0   &   0.3 /  0.3  \\ 
  3665  &   1.8 /  1.7   &   2.1 /  1.9   &   2.2 /  2.0   &   2.4 /  2.0  \\ 
  3685  &   2.8 /  2.7   &   3.1 /  2.8   &   3.2 /  2.8   &   3.2 /  2.7  \\ 
  3705  &   0.7 /  0.8   &   1.1 /  1.0   &   1.5 /  1.4   &   1.7 /  1.5  \\ 
  3725  &   0.5 /  0.5   &   0.8 /  0.8   &   0.9 /  0.9   &   0.9 /  0.9  \\ 
  3745  &   0.0 /  0.0   &   0.0 /  0.0   &   0.0 /  0.0   &   0.0 /  0.0  \\ 
  3765  &   0.1 /  0.2   &   0.0 /  0.0   &   0.0 /  0.0   &   0.0 /  0.0  \\ 
  3785  &   0.0 /  0.0   &   0.0 /  0.0   &   0.0 /  0.0   &   0.0 /  0.0  \\ 
  3805  &   0.0 /  0.0   &   0.0 /  0.0   &   0.0 /  0.0   &   0.0 /  0.0  \\ 
\end{tabular}
\end{center}
\end{table*}
\renewcommand{\arraystretch}{1.0}

%
%
\section{\label{sec:Syst} Systematic uncertainties}
Various sources of systematic uncertainty have been investigated for the cross section measurements. The uncertainty on the integrated luminosity $\cal{L}$ determined using Bhabha events is estimated to be 1.0\,\%~\cite{bes3_lumi_paper}. The uncertainties of the $\eta_{\rm c}$ resonance parameters $(m_0,\Gamma_0)$ for the signal line shape as well as those on the branching ratios are taken from the PDG~\cite{PDG2016}. There are uncertainties introduced by the {\mbox BESIII} detector, due to knowledge of the charged particle and neutral reconstruction, as well as the PID. For charged tracks, single $\gamma$'s and reconstructed $\pi^0$'s, we apply an error of 1\,\% each, for reconstructed $K_S^0$ 1.2\,\% per particle, and as PID uncertainty, we apply 1\,\% per identified charged track, following earlier studies~\cite{bes3_track_syst,bes3_phot_syst,hcpipi_BESIII}. There are uncertainties associated with reconstruction efficiencies for each $\eta_{\rm c}$ decay mode, as well as with the radiative-correction factors $\kappa_{i}$ due to the used resonance line shape (Sec.\,\ref{subsec:RadCorr}). Moreover, there is also an uncertainty in the background description using analytical functions, namely polynomials of $2^{\rm nd}$, $3^{\rm rd}$ and $4^{\rm th}$ order. Since we checked and corrected for possible biases introduced by the simultaneous fits, there is also a connected uncertainty taken into account.  
 
For these systematic uncertainties (sources a) -- i) in Tab.\,\ref{tab:RankingSystError_Etac_M8_4230}), the corresponding parameters have randomly been varied (drawn from a Gaussian distribution) at the same time within the given uncertainties, and the simultaneous fits have been repeated many times for each production channel of $\eta_{\rm c}$ or $Z_{\rm c}$ plus recoil particles. To take into account the systematic uncertainty introduced by the background descriptions using polynomials, we also randomly vary the polynomial order for each case in the fit repetitions. The root-mean-square of the resultant fitted cross sections is assigned as the corresponding systematic uncertainty. By this procedure of random variation of parameters and refitting, possible correlations between the different parameters are taken into account.
\begin{table}[h!]
\begin{center}
  \caption{\label{tab:RankingSystError_Etac_M8_4230} 
    List of systematic uncertainties separated for the individual sources a) -- i) ($\Delta \sigma_{\rm rdm-vary}$) 
    and j) -- l) quoted separately for the example of the channel $\eta_{\rm c}\pi^+ \pi^- \pi^0$ at $\sqrt{s}=4.23$\,GeV. 
    Those contributions that do not enter into the computation of the signal significances are marked by (*).   
  }
\begin{tabular}{l|c}
  source  &  uncertainty [\%]  \\ \hline 
{a)}~~branching fractions (*)                &   8.7  \\ 
{b)}~~tracking/neutrals reco./PID (*)        &   7.2  \\ 
{c)}~~ISR correction(*)                      &   5.8  \\ 
{d)}~~integrated luminosity (*)              &   1.0  \\ 
{e)}~~reconstruction efficiencies (*)        &   0.8  \\ 
{f)}~~fit range variation                        &   6.1  \\ 
{g)}~~background shape                       &   5.9  \\ 
{h)}~~$\eta_{\rm c}$ parameters                    &   1.4  \\ 
{i)}\;~~bias correction                      &   0.7  \\ \hline 
 $\Delta \sigma_{\rm rdm-vary}$      &  13.8  \\ \hline
{j)}~~kinematic fit $\chi^2$ cut         &   4.4  \\
{k)}~vertex fit $\chi^2$ cut            &   1.3  \\
{l)}~~veto cuts                          &   0.3  \\ \hline
  $\Delta \sigma_{\rm sys, total}$         &  14.3  \\
\end{tabular}
\end{center}
\end{table}

The uncertainties introduced by the selection criteria of the two $\chi^2$ values from the kinematic and the vertex fit, and those by the veto cuts (sources j) -- l) in Tab.\,\ref{tab:RankingSystError_Etac_M8_4230}) are estimated by varying each of the cuts by $\pm 10\,\%$ and repeating the simultaneous fits. The standard deviation of the three fit results is taken as an estimate for the systematic uncertainty introduced by these selection and veto cuts.

The final total systematic uncertainty $\sigma_{\rm sys,total}$ is then computed as the quadratic sum of these three contributions from sources j) -- l) and the systematic uncertainty $\sigma_{\rm rdm-vary}$ obtained for sources a) -- i) from the simultaneous random variation.
 
For the given example in Tab.\,\ref{tab:RankingSystError_Etac_M8_4230}, the contribution to the systematic uncertainty from sources j) -- l) varies from 4.4\,\% down to 0.3\,\%, and adding them in quadrature to $\Delta \sigma_{\rm rdm-vary} = 13.8$\,\% results in an total systematic uncertainty of $\Delta \sigma_{\rm sys, total} = 14.3$\,\%.

For the upper limits on the Born cross sections $\sigma_{\rm Born}$, the systematic uncertainties are taken into account by convolving the likelihood distributions (Fig.\,\ref{fig:NegLogLikelihood_etac_M8_4230}) with a Gaussian of standard deviation $\sigma_{\rm sys,total}$, which broadens them (blue solid curve in Fig.\,\ref{fig:NegLogLikelihood_etac_M8_4230}), leading to correspondingly more conservative UL$_{90}$ values. 

To take into account the systematic uncertainties in the signal significances $\cal{S_{\rm tot}}$, again the likelihood distributions are convolved with a Gaussian with width set to the standard deviation that corresponds to the relevant systematic uncertainty. 

There are two groups of sources of systematic uncertainties, namely those which affect the observed event yield returned by the simultaneous fit directly, and those which do not. The latter group affects the fit result merely via a multiplicative factor entering Eq.\,(\ref{Eq.simuFit_ISR}). Therefore, the systematic-uncertainty contributions introduced by  sources a) -- e) in Tab.\,\ref{tab:RankingSystError_Etac_M8_4230}) are not relevant when calculating the signal significances, under which possible $\eta_{\rm c}$ (or $Z_{\rm c}$) production cross sections are observed and reported. For the calculation of the UL$_{\rm 90}$ values,  the total systematic uncertainty $\sigma_{\rm sys,total}$ is taken into account.

When fitting a line shape to the measured Born cross section (Fig.\,\ref{fig:EtacResults_etac_M8}), only those systematic uncertainties have to be taken into account that are uncorrelated between the different center-of-mass energies, since only those affect the absolute cross section measurements differently at each value of $\sqrt{s}$. Therefore, the sources of systematic uncertainties a), b) and d) are omitted for our fit of the energy-dependent line shape (Fig.\,\ref{fig:EtacResults_etac_M8}), but not the sources c) and e), since the uncertainties introduced by the ISR correction and the reconstruction efficiencies have a $\sqrt{s}$ dependence.
 
%
%
\section{\label{sec:Summary} Summary of results}
In summary, we have studied $\eta_{\rm c} + {\rm light\,recoil} $ production in the range of $\sqrt{s}$ from 4.18 to 4.60\,GeV in the three exclusive channels $e^+e^-\to\eta_{\rm c}\pi^+\pi^-\pi^0$, $e^+e^-\to\eta_{\rm c}\pi^+\pi^-$ and $e^+e^-\to \eta_{\rm c}\pi^0\gamma$.

The process $e^+e^-\to \eta_{\rm c}\pi^+\pi^-\pi^0$ is observed for the first time with an energy-dependent Born cross section measured to be in agreement with the hypothesis of the production of an intermediate $Y(4260)$ resonance decaying to the $\eta_{\rm c}\pi^+\pi^-\pi^0$ final state. The largest observed cross section of $\sigma_{\rm Born}= 46.1^{+9.5}_{-9.4}\pm 6.6$\,pb as measured at $\sqrt{s}=4.23$\,GeV has also the highest  significance value of 4.1$\sigma$ (5.1$\sigma$), when including (not including) systematic uncertainties. Summing up all six cross section values at the different $\sqrt{s}$ points, including uncertainties, results in a simple (value over uncertainty) significance of $5.2\,\sigma$ ($5.9\,\sigma$) when including (not including) the systematic uncertainties. This result will be valuable in helping to understand the nature of the $Y(4260)$. The cross sections of the other two $\eta_{\rm c}$ production modes with recoil particles $\pi^+\pi^-$ and $\pi^0\gamma$ are found to be consistent with zero and corresponding upper limits are provided.

Based on the data at $4.23$\,GeV, we have performed a width and mass scan to search for possible charged and neutral $Z_{\rm c}$-like $\eta_{\rm c}\pi$ resonances in $\eta_{\rm c}\pi^+\pi^-\pi^0$ final states. While no signal is found for a charged $Z_{\rm c}^\pm \to \eta_{\rm c}\pi^\pm$ state, the measured cross sections are found to be more significant for a neutral $Z_{\rm c}^0 \to \eta_{\rm c}\pi^0$ state. Here, we find a maximum signal significance of 2.8$\sigma$ (3.2$\sigma$) when including (not including)  systematic uncertainties. This evidence disappears, however, when applying a statistical penalty of about $1.2\sigma$ due to the ``Look-elsewhere'' effect, according to which the highest observed significance of $3.2\sigma$ for the assumed resonance parameters of $(m_{\rm Z_{\rm c}}, \Gamma_{\rm Z_{\rm c}})=(3685\,{\rm MeV}/c^2, 28\,{\rm MeV})$ is reduced down to about $2\sigma$.

The di-quark model that predicts a $Z_{\rm c}$ candidate with $J^P= 0^+$ beneath the open charm threshold is presently not supported by the BESIII data.  A future analysis based on a larger data set would reveal whether the observed $2\sigma$ effect is a hint for such a state or not~\cite{Ablikim:2019hff}. Once also higher energy data has been accumulated above 4.6\,GeV with BESIII, the reported evidence for a $Z_{\rm c}(4096)$ by the LHCb Collaboration can be checked in our data.

\begin{acknowledgments}
The BESIII collaboration thanks the staff of BEPCII and the IHEP computing center for their strong support. This work is supported in part by National Key Basic Research Program of China under Contract No. 2015CB856700; National Natural Science Foundation of China (NSFC) under Contracts Nos. 11625523, 11635010, 11735014, 11822506, 11835012, 11935015, 11935016, 11935018, 11961141012; the Chinese Academy of Sciences (CAS) Large-Scale Scientific Facility Program; Joint Large-Scale Scientific Facility Funds of the NSFC and CAS under Contracts Nos. U1732263, U1832207; CAS Key Research Program of Frontier Sciences under Contracts Nos. QYZDJ-SSW-SLH003, QYZDJ-SSW-SLH040; 100 Talents Program of CAS; INPAC and Shanghai Key Laboratory for Particle Physics and Cosmology; ERC under Contract No. 758462; German Research Foundation DFG under Contracts Nos. Collaborative Research Center CRC 1044, FOR 2359; Istituto Nazionale di Fisica Nucleare, Italy; Ministry of Development of Turkey under Contract No. DPT2006K-120470; National Science and Technology fund; STFC (United Kingdom); The Knut and Alice Wallenberg Foundation (Sweden) under Contract No. 2016.0157; The Royal Society, UK under Contracts Nos. DH140054, DH160214; The Swedish Research Council; U. S. Department of Energy under Contracts Nos. DE-FG02-05ER41374, DE-SC-0012069.
\end{acknowledgments}

\clearpage

\end{document}